\documentclass[10pt,aps,pra,twocolumn,superscriptaddress,floatfix,nofootinbib,showpacs,longbibliography]{revtex4-1}
\pdfoutput=1

\usepackage[utf8]{inputenc}  
\usepackage[T1]{fontenc}     
\usepackage[british]{babel}  
\usepackage[scaled=0.86]{berasans} 
\usepackage[colorlinks=true, 
            citecolor=blue, 
            urlcolor=blue,
            hyperindex,
            breaklinks]{hyperref}  
\usepackage{graphicx}
\usepackage[babel]{microtype}  
\usepackage{amsmath,amssymb,amsthm,bm,amsfonts,mathrsfs,bbm} 

\usepackage{algorithm}
\usepackage{algpseudocode}

\usepackage{physics}
\usepackage{xcolor}

\usepackage{tikz}
\usetikzlibrary{quantikz2}

\usepackage{svg}

\usepackage{dcolumn}

\theoremstyle{plain}
\newtheorem{theorem}{Theorem}
\newtheorem{lemma}{Lemma}

\newtheorem{definition}{Definition}

\begin{document}


\title{Efficient graph-diagonal characterization of noisy states distributed over quantum networks via Bell sampling}

\author{Zherui Jerry Wang}%
\thanks{These authors contributed equally}
\affiliation{Networked Quantum Devices Unit, Okinawa Institute of Science and Technology Graduate University, Okinawa, Japan}%
\affiliation{$\langle aQa^L \rangle$ Applied Quantum Algorithms, Universiteit Leiden, the Netherlands}%
\affiliation{Instituut-Lorentz, Universiteit Leiden, P.O. Box 9506, 2300 RA Leiden, The Netherlands}
\affiliation{QuTech, Delft University of Technology, Lorentzweg 1, 2628 CJ Delft, The Netherlands}
\author{Joshua Carlo A. Casapao}%
\thanks{These authors contributed equally}
\affiliation{Networked Quantum Devices Unit, Okinawa Institute of Science and Technology Graduate University, Okinawa, Japan}%
\author{Naphan Benchasattabuse}%
\thanks{These authors contributed equally}
\affiliation{Keio University Shonan Fujisawa Campus, 5322 Endo, Fujisawa, Kanagawa 252-0882, Japan}%
\author{Ananda G. Maity}%
\affiliation{Networked Quantum Devices Unit, Okinawa Institute of Science and Technology Graduate University, Okinawa, Japan}%
\affiliation{School of Physical Sciences, Indian Institute of Technology Goa, Ponda, Goa 403401, India}%
\author{Jordi Tura}%
\affiliation{$\langle aQa^L \rangle$ Applied Quantum Algorithms, Universiteit Leiden, the Netherlands}%
\affiliation{Instituut-Lorentz, Universiteit Leiden, P.O. Box 9506, 2300 RA Leiden, The Netherlands}
\author{Akihito Soeda}%
\affiliation{Principles of Informatics Research Division, National Institute of Informatics, 2-1-2 Hitotsubashi, Chiyoda-ku, Tokyo, 101-8430, Japan}%
\affiliation{Department of Informatics, School of Multidisciplinary Sciences, SOKENDAI, 2-1-2 Hitotsubashi, Chiyoda-ku, Tokyo 101-8430, Japan}%
\affiliation{Department of Physics, Graduate School of Science,
The University of Tokyo, 7-3-1 Hongo, Bunkyo-ku, Tokyo 113-0033, Japan}%
\author{Michal Hajdu\v{s}ek}%
\affiliation{Keio University Shonan Fujisawa Campus, 5322 Endo, Fujisawa, Kanagawa 252-0882, Japan}%
\author{Rodney Van Meter}%
\affiliation{Keio University Shonan Fujisawa Campus, 5322 Endo, Fujisawa, Kanagawa 252-0882, Japan}%
\author{David Elkouss}%
\affiliation{Networked Quantum Devices Unit, Okinawa Institute of Science and Technology Graduate University, Okinawa, Japan}%

\begin{abstract}
Graph states are an important class of entangled states that serve as a key resource for distributed information processing and communication in quantum networks.
In this work, we propose a protocol that utilizes a Bell sampling subroutine to characterize the diagonal elements in the graph basis of noisy graph states distributed across a network.
Our approach offers significant advantages over direct diagonal estimation using unentangled single-qubit measurements in terms of scalability.
Specifically, we prove that estimating the full vector of diagonal elements requires a sample complexity that scales linearly with the number of qubits ($\mathcal{O}(n)$), providing an exponential reduction in resource overhead compared to the best known $\mathcal{O}(2^n)$ scaling of direct estimation. 
Furthermore, we demonstrate that global properties, such as state fidelity, can be estimated with a sample complexity independent of the network size.
Finally, we present numerical results indicating that the estimation in practice is more efficient than the derived theoretical bounds.
Our work thus establishes a promising technique for efficiently estimating noisy graph states in large networks under realistic experimental conditions.
\end{abstract}

\maketitle

\section{Introduction\label{intro}}
Significant efforts have been devoted to the development of quantum networks, in which clients are capable of performing distributed information processing and secure communication tasks mediated by quantum mechanics \cite{Buhrman03,Broadbent09,Gottesman12,Kmr14,Fitzsimons17,IloOkeke18,Cuomo20,Bauml20}.
The construction of these quantum networks requires the distribution of long-range, high-quality entangled resource states to target applications.
Well-studied approaches to establish long-range entanglement utilize quantum repeaters that perform entanglement swapping on shorter bipartite entangled resources between network clients \cite{Muralidharan2016,Goodenough21,azuma2023quantum}.
However, schemes that instead use on-demand multipartite entanglement have recently been proposed, offering a more flexible and highly modular architecture for future quantum networks
\cite{Pirker18,Pirker2019stack}.

Among the multipartite resource states that we can utilize are graph states. Graph states are a particularly useful class of states that serve as fundamental primitives not only for many network applications, but also as building blocks for more complex quantum states~\cite{Zwerger2014,Pirker18,Hahn2019,Matsuzaki10,Cuquet12,Cullen22,Fischer21,Meignant19,Epping16,MiguelRamiro2023,fan2024}.
In these states, physical systems described by qubits are represented by the nodes of a graph, whereas the entanglement between them is represented by the edge set.
Due to their broad applicability in distributed information processing \cite{Hein2006entanglement,Bell2014,Azuma2015,Shettell20}, significant theoretical \cite{Brown12,Inaba14,Monroe14} and experimental \cite{Lu2007experimental,Tokunaga08,Yao2012,Wang16,Wang18,Gong19,Thomas2022efficient} efforts have focused on the direct generation of increasingly large graph states.

Unfortunately, robustly preparing and manipulating graph states that span over a large portion of a network can be both resource-intensive and time-consuming. 
Operational errors due to intrinsic noise that affect network devices degrade the quality of any distributed entanglement.
These errors effectively reduce network performance in measures such as throughput and latency, or even completely hinder the network in meeting quality thresholds for applications \cite{azuma2023quantum,Azuma21,Chakraborty20,Goodenough21,Coopmans21,Satoh22}. 
As such, having sufficient information about the quality of distributed resource states allows us to intelligently tackle this unavoidable noise by using quantum error correction \cite{Zwerger2014,Muralidharan2016} and multipartite entanglement distillation \cite{Dur2003Multiparticle,Kruszynska2006entanglement}, for example.
Therefore, protocols that perform estimation and verification with reliable accuracy and precision are indispensable as on-demand quality assurance tools \cite{Eisert2020}.  

Several estimation and verification tools already exist in the literature, including self-testing \cite{mckague2014selftesting,Supic2020}, quantum state tomography \cite{bisio2009quantum}, network tomography \cite{Andrade22,Andrade23,Andrade24}, fidelity-based entanglement witnesses \cite{Bourennane2004,Guhne2009entanglement}, randomized measurement-based verification protocols \cite{Emerson05,Knill08,Dankert09,Magesan11,Magesan12,Wallman16,Erhard19,Hashim21,Helsen22}, and distillation-based estimation \cite{maitynoise,casapao2024disti,casapao2025doubles}.
However, given resource constraints and device limitations, quantum networks must balance information gain, implementation complexity, and the strength of the assumptions required \cite{Eisert2020}. 
Such trade-offs are especially relevant for characterizing a graph state, where the estimation efficiency can be seriously compromised by the complexity of the graph. 
In particular, we note that estimation tools can become impractical because the dimensionality of the unknown parameter space of interest scales unfavorably as the system size increases.

This scalability problem is no more pronounced than in quantum state tomography. 
Tomography is a standard approach to estimate the density matrices of multipartite entangled resources generated experimentally \cite{Haffner2005,Chen2025efficient}.
This approach relies on fully trusted and well-characterized measurement devices, along with precise knowledge of the system's dimensions.
However, in terms of sampling complexity, a complete tomographic reconstruction of an unknown $n$-qubit state up to some additive error $\varepsilon$ in terms of the trace distance requires $\Omega(4^{n}/\varepsilon^2)$ samples with collective (that is, unrestricted) measurements and $\Omega(8^{n}/\varepsilon^2)$ samples with only independent (product or unentangled) measurements \cite{Haah2017sample}.
This exponential dependence on $n$ also translates into a large number of tomographic base measurements.

Optimized tomographic strategies can be devised by leveraging prior information, ancillary qubits, or oracle primitives, particularly when only partial information is required or specific state properties are known.
From a hardware perspective, probing states via ancilla qubits with tunable interactions has proven effective in minimizing experimental complexity; for instance, full-state tomography of Rydberg atom graph states was recently demonstrated using a reconfigurable ancillary atom system scaling up to 26 atoms~\cite{Kim23}.
In the theoretical domain of learning unknown graph structures through quantum queries, it has been established that $\Theta(n)$ samples are necessary and sufficient, yet this sample complexity improves if structural information---such as the maximum degree $k$---is available~\cite{montanaro2021}.
An effective query model in this context employs a Bell sampling protocol~\cite{montanaro2017learning}, where corresponding qubits from two simultaneous copies of the pure graph state are measured in the Bell basis.
Even when restricted to product measurements (single-qubit $X$ and $Z$ bases) on a graph that satisfies $n\gg k$, only $\mathcal{O}(k\log (1/\delta^2)+k^2\log n)$ samples are required to efficiently learn the graph with probability $1 - \delta$~\cite{Ouyang22}.
In scenarios where the target graph is already defined, stabilizer measurements can be used to certify graph states in quantum networks for specific applications~\cite{Markham20}.
For fidelity estimation tasks, a small number of stabilizer measurements selected by importance weighting can minimize resource costs~\cite{flammia2011direct}.
Furthermore, time-efficient estimation algorithms have been developed for states satisfying specific conditions, including Bell difference sampling for pure states close to some stabilizer states~\cite{Grewal2024improved}, stabilizer bootstrapping for polynomial-time verification~\cite{Chen2025}, and single-qubit fidelity estimation under depolarizing noise models~\cite{tanizawa2023}.

Beyond these property-specific protocols, there are also many situations where complete tomographic knowledge is unnecessary, and the objective is simply to estimate the expectation values of a set of observables within a small error and with high probability.
This is known as the shadow tomography problem~\cite{aaronson2018}, and early results demonstrated that probing the state with fewer measurements and samples than full tomography suffices~\cite{aaronson2018, brandaoQuantumSDPSolvers2019}.
A significant advancement in this area was the development of classical shadows~\cite{Huang2020}---approximate classical descriptions of quantum states.
Crucially, this strategy allows for the simultaneous prediction of numerous properties without requiring prior knowledge of the specific observables to be estimated.
Consequently, the measurement overhead scales only logarithmically with the number of properties considered.
Recently, this framework has been further optimized to reduce circuit depth; for instance, a resource-efficient scheme using equatorial stabilizer measurements was proposed in~\cite{Park2025}, achieving low sampling complexity for estimating a large class of observables in $n$-qubit systems.

Alternatively, one may seek to estimate global properties of the density matrix---such as its spectrum---which provides information about entanglement and fidelity without requiring explicit full-state reconstruction.
Although spectrum estimation was presumed to be as resource-intensive as full tomography, recent results demonstrate that using only unentangled measurements, it is possible to construct algorithms that require fewer samples than standard tomography; specifically, $\mathcal{O}(8^n(\log(n)/n)^4/\varepsilon^6)$ samples suffice for an $n$-qubit state~\cite{pelecanos2025beating}.
Regarding entangled measurements, while it is conjectured that they could offer improved efficiency, the extent of this advantage remains unclear.
Currently, the best known lower bound is $\Omega(2^n/\varepsilon^2)$, derived from the maximally mixed state testing~\cite{ODonnell2015quantum}.
However, recent numerical evidence suggests this bound is not tight and that the improvement over unentangled measurements is likely only sub-polynomial, implying that the true sample complexity for spectrum estimation with entangled measurements likely remains comparable to full tomography~\cite{pelecanos2025beating}.

In this work, we describe Bell Sampling for Quantum Networks (BSQN) as a means of reducing the resource overhead of characterizing noisy graph states distributed over a network by foregoing a full tomographic reconstruction and performing a partial characterization instead.
In our BSQN protocol, we use a Bell sampling protocol as a subroutine, measuring the stabilizer expectation values given two copies of the noisy state. 
The remainder of the BSQN protocol then post-processes the resulting statistics to estimate all the state's diagonal elements in the graph-state basis. 
These diagonal elements provide information on the noisy $n$-partite Pauli channel during distribution, highlighting the value of our protocol as an error detection tool. 
Although BSQN assumes trusted devices on the nodes to perform measurements, our protocol is nonetheless sample-efficient and less experimentally demanding in the near term compared with device-independent methods. 

We show that $\mathcal{O}\!\left( {(n+\log(1/\delta))}/{\varepsilon^4} \right)$ samples of the $n$-qubit state are sufficient to successfully estimate the diagonal elements when the fidelity is greater than $1/2$. 
That is, the sample complexity grows at most linearly with the size of the system. 
Moreover, we show that $\mathcal{O}\!\left({\log(1/\varepsilon^2\delta)}/{\varepsilon^4}\right)$ samples are sufficient to estimate fidelity with high probability, regardless of the size of the system. 
Our results demonstrate better performance than a direct estimation of the diagonal elements using single copies. 

The remainder of the paper is organized as follows. 
In Section~\ref{sec:preliminaries}, we introduce the graph state formalism and outline the Bell sampling protocol. 
In Section~\ref{sec:estimating}, we present our main results for characterizing noisy graph states distributed over a network using BSQN and discuss the relevant sample complexity. 
Notably, we demonstrate that BSQN becomes more efficient compared to a direct graph-diagonal estimation (DGE) for a larger number of qubits. 
In Section~\ref{sec:errordetection}, we discuss BSQN as a Pauli error detection tool.
In Section~\ref{sec:results}, we elaborate on our findings through numerical experiments for explicit examples and provide an analysis of how the protocol scales with the number of qubits. 
Finally, in Section \ref{sec:summary}, we conclude with a discussion of potential future extensions of our work.


\section{Preliminaries}\label{sec:preliminaries}

Let us now introduce some basic notation and essential tools that form the foundation of this work. 
We first recall the mathematical formalism of graph states and then follow this with an overview of the Bell sampling protocol which we employ for our estimator. 
For this work, we define the $n$-Pauli group as $\mathcal{P}_n$, and the set $\mathcal{P}_n$ modulo phases as
$\mathcal{P}_n^+ \equiv \{I,X,Y,Z\}^{\otimes n}$, with size $\abs{\mathcal{P}_n^+} = 4^n$. 

\subsection{Graph states}\label{ss:graph_state}

Graph states \cite{hein2004multiparty,Hein2006entanglement} are a class of multipartite entangled states represented by (undirected and simple) graphs $G=(V,E)$, where each vertex belonging to the set $V$ corresponds to a physical qubit, and each edge belonging to the set $E$ represents an entangled link. 
Each qubit is initially prepared in the $\ket{+}$ state, and then a link is created through a controlled-$Z$ operation $CZ_{u,v}$ (defined as $\ket{0}\bra{0}_u\otimes \mathbb{I}_v+\ket{1}\bra{1}_u\otimes Z_v$) if two vertices $u,v$ share a graph edge. 
Formally, we have the following definitions.
\begin{definition}[Graph states\label{def:graph_state}]
    A graph state $\ket{\Phi_G}$ associated with a graph $G=(V,E)$ is defined as
    \begin{equation}
        \ket{\Phi_G} \equiv \prod_{(u,v) \in E} CZ_{u,v} \ket{+}^{\otimes \abs{V}}.
    \end{equation}
    This is a simultaneous $+1$ eigenstate of the set of mutually commuting stabilizer generators $\{S_v\}_{v\in V}$, each defined as
    \begin{equation}\label{eq:graph_state_stabilizer}
        S_v \equiv X_v  \prod_{u \in \mathrm{NN}(v)} Z_u,
    \end{equation}
    where $\mathrm{NN}(v)$ denotes the set of vertices adjacent to the vertex $v$.
\end{definition}
\noindent Let $n=\abs{V}$ denote the total number of physical qubits that describe $\ket{\Phi_G}$ ($n$ is also called the graph order). 
With the generators defined above, we denote the $n$-qubit stabilizer group corresponding to $\ket{\Phi_G}$ as $\mathcal{S}_G \equiv \langle S_1, S_2, \cdots, S_n \rangle$, with $\abs{\mathcal{S}_G} = 2^n$. 
We further define $\mathcal{S}_G^+$ as the set $\mathcal{S}_G$ modulo phases,
a notation introduced because global phases can be ignored in our later calculations.
\begin{definition}[Graph-state basis\label{def:graph_basis}]
    The graph-state basis $\mathcal{G}$ associated with the graph state $\ket{\Phi_G}$ is a complete and orthonormal basis of the Hilbert space of $n$ qubits defined as 
    \begin{equation}
        \mathcal{G} \equiv 
        \left\{ \ket{\Phi_b}: \ket{\Phi_b}=\left(\bigotimes_{i=1}^{n} Z^{b_i} \right)\ket{\Phi_G}  \right\}_{ b \in \mathbb{F}_2^{n}},
    \end{equation}
where $b \equiv \overline{b_1 \cdots b_{1<i<n} \cdots b_{n}}\in \mathbb{F}_2^{n} \equiv \{0,1\}^n$ 
is an $n$-bit string.
\end{definition}
\noindent We also denote that $\ket{\Phi_G}$ can be rewritten as $\ket{\Phi_{0}}$.

Graph states are suitable for quantum communication and computing networks, in which the entanglement between physical devices within a network is naturally encoded by the topology of the underlying graph \cite{hein2004multiparty}.  
Moreover, graph states are known to be locally unitarily equivalent to generic stabilizer states, and the actions of Pauli noise on graph states can be efficiently tracked \cite{mor-ruiz2023noisy}.

\subsection{Bell sampling}\label{ss:bell_sampling}

We now describe the Bell sampling protocol. Bell sampling \cite{montanaro2017learning,hangleiter2023bell} employs two identical copies of an unknown state, performing transversal Bell basis measurements on all corresponding qubit pairs across the bipartition of the two copies. 
\begin{definition}[Bell sampling protocol\label{def:bell_sampling}]
    Let $\rho$ be an $n$-qubit state. The Bell sampling protocol proceeds as follows:
    \begin{enumerate}
        \item Prepare $\rho \otimes \rho$.
        \item Apply a qubit-wise CNOT operation between the first and second copies, using the qubits of the first copy as the controls and the qubits of the second copy as the targets.
        \item Apply a Hadamard gate to each qubit in the first copy.
        \item Measure both copies qubit-wise in the computational basis, returning a $2n$-bit string.
\end{enumerate}
\end{definition}
This protocol is known to efficiently estimate a stabilizer state $\rho$ from the measurement statistics of $\mathcal{O}(n)$ pairs of $\rho$ up to a constant error \cite{montanaro2017learning}. 
Moreover, the same measurement statistics can reveal many properties of the quantum circuit that prepare $\rho$~\cite{hangleiter2023bell}. 
Other recent applications of Bell sampling techniques include learning stabilizer entropies \cite{haug2023efficient}, testing stabilizerness \cite{gross2021schur}, quantifying the magic resource for pure quantum states \cite{haug2023scalable}, assessing purity of a quantum state \cite{hangleiter2023bell}, and establishing an exponential advantage of quantum machine learning over its classical counterpart \cite{huang2021information}. 
In particular, Bell sampling can be used as a subroutine for Pauli shadow tomography to efficiently and simultaneously estimate expectation values of non-commuting $n$-Pauli operators \cite{huang2021information}. 
The key insight is that given $P,P'\in\mathcal{P}_n$, where $P$ may not commute with $P'$ and therefore is not necessarily simultaneously measurable, the commutation relation $[P \otimes P, P' \otimes P']=0$ always holds. 
The transverse Bell measurements over two copies of $\rho$ are then used to reveal the quantities $\abs{\Tr(\rho P)}^2 = \Tr((\rho \otimes \rho)(P \otimes P))$.

In estimating multi-qubit states, we recall that any $n$-qubit density matrix $\rho$ can be written as
\begin{equation}
    \rho = \frac{1}{2^n}\sum_{P\in \mathcal{P}_n^+}\Tr(\rho P)P.
\end{equation}
We then see that the learning task using Pauli shadow tomography consists of two distinct phases. 
The first phase focuses on estimating $\abs{\Tr(\rho P)}^2$ via Bell sampling, which produces the desired magnitudes $\abs{\Tr(\rho P)}$. 
The second phase aims to determine the sign of $\Tr(\rho P)$, which is generally not straightforward. 
It was shown that the magnitude of the expectation value of any $P$ can be estimated with $\mathcal{O}(1/\varepsilon^4)$ samples using Bell sampling, but requires an additional $\mathcal{O}(n/\varepsilon^2)$ samples to be stored in a quantum memory to estimate the corresponding sign by majority voting (which is unrelated to Bell sampling), where $\varepsilon$ is the additive error \cite{huang2021information}.
We note that this majority voting jointly measures the signs of all $4^n$ Pauli operators in $\rho$. 
In contrast, an optimal Pauli shadow tomography strategy using only single-copy measurements can achieve a sample complexity that scales like $1/\varepsilon^2$ but is exponentially dependent on $n$ \cite{chen2024optimal}.

\section{Estimating graph states over a network} \label{sec:estimating}

\begin{figure}[t]
    \centering
    \includegraphics[width=1.0\linewidth]{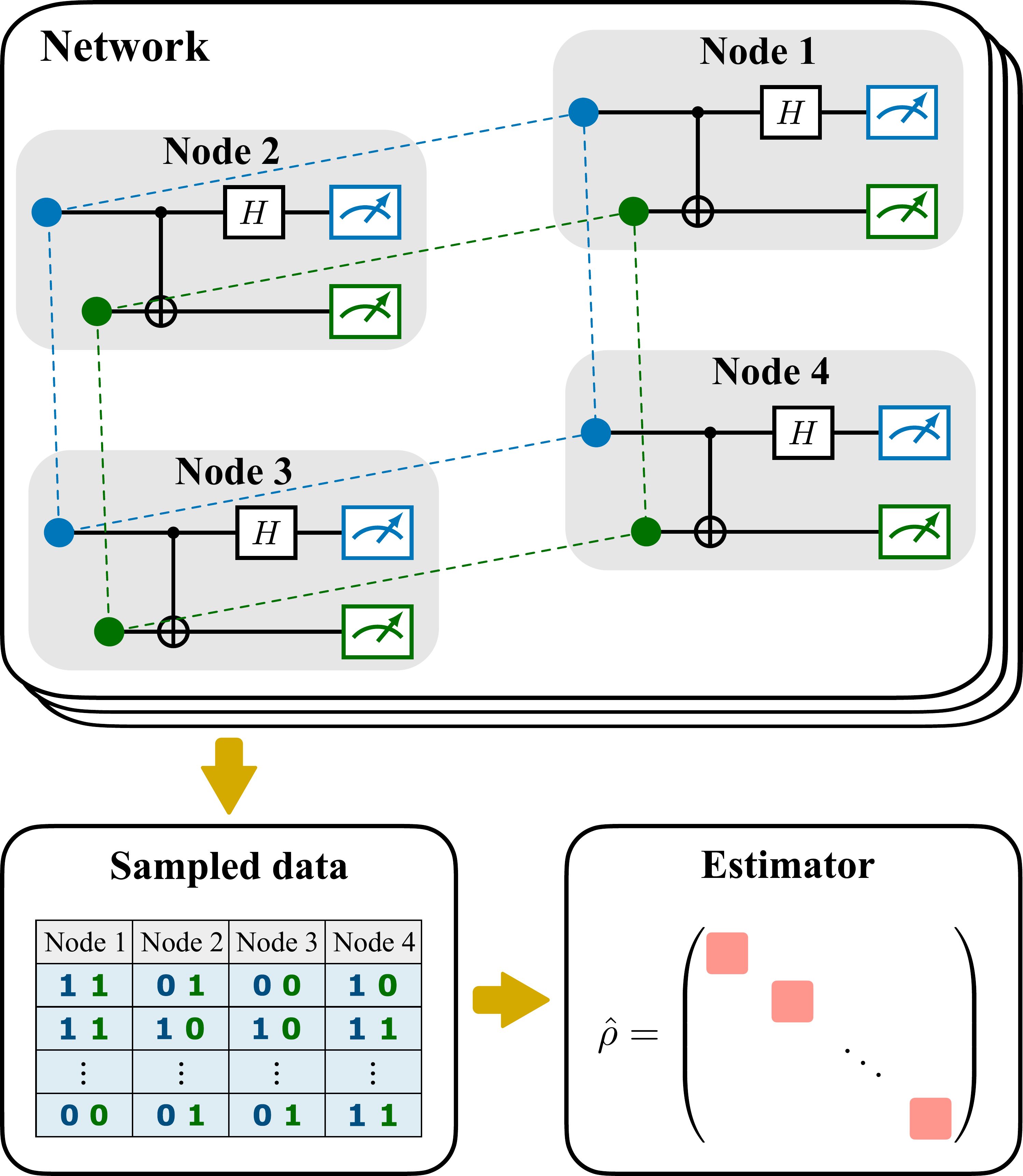}
    \caption{\label{fig:protocol}
    \textbf{Bell Sampling for Quantum Networks (BSQN).} Two copies of the noisy graph state are distributed across the network nodes. The initial stage of BSQN is a Bell sampling subroutine: each node performs a qubit-wise CNOT operation, followed by applying a Hadamard gate to the control qubit and measuring in the computational basis. 
    The measurement statistics are then collected and post-processed to estimate the diagonal elements of the distributed noisy graph state.
    }
\end{figure}

In this section, we introduce the BSQN protocol. Our protocol performs partial characterization of noisy graph states distributed over a quantum network by utilizing a Bell sampling subroutine.
Figure~\ref{fig:protocol} provides a visualization of the protocol, while Algorithm~\ref{alg} provides a summary of the protocol.

When distributing a target graph state $\ket{\Phi_G}$ across the network, the inevitable presence of noise transforms $\ket{\Phi_G}$ into a noisy state $\rho = \mathcal{N}(\Phi_{G})$, where $\mathcal{N}$ is some completely positive trace-preserving map. 
Assuming that the actual states $\rho$ are independent and identically distributed (i.i.d.) and the resulting fidelity $F(\Phi_{G}, \rho) > {1}/{2}$, our protocol efficiently estimates all $2^n$ diagonal elements of $\rho$ in the graph basis $\mathcal{G}$. 
We show that the expectation value $\Tr(\rho P)$, with $P\in \mathcal{S}_G^+$, can be efficiently learned from the measurement statistics of the Bell sampling protocol via additional classical post-processing. 
Our protocol also removes the need for quantum memory in estimating the signs of the expectation values, since these signs can be classically determined from the measurement statistics. 
From an implementation perspective, our protocol benefits from the fact that Bell sampling measurements are device-local and that the measurement circuits in each node are fixed. 

\begin{algorithm}[H]
\caption{: Bell Sampling for Quantum Networks (BSQN)}\label{alg}
\begin{algorithmic}[1]
\Require $N$ pairs of state $\rho \otimes \rho$, in which $\rho$ has fidelity $F(\Phi_{G}, \rho) > {1}/{2}$.
\Ensure Estimation $\hat{a}_b$ of the diagonal elements $a_b = \bra{\Phi_b}\rho\ket{\Phi_b}$ for all $b\in \mathbb{F}_2^n$.
\State Perform Bell sampling measurements on pairs $\rho \otimes \rho$ and send the results to the verifier.
\State Verifier calculates an estimation for $|\Tr(\rho P)|^2$ for each $P \in \mathcal{S}_G^+$ (see Appendix \ref{app:theorem-sample-complexity}), and obtains both the magnitude $|\Tr(\rho P)|$ and the sign i.e., $\mathrm{sgn}(\Tr(\rho P))$.
\Statex \Comment{Estimation can be also performed with few randomly sampled stabilizer elements (see Appendix \ref{app:random_sampling})}
\State Obtain $\hat{a}_b$.
\end{algorithmic}
\end{algorithm}
For the rest of the work, we let $\mathbf{a}$ be the real vector of diagonal entries of $\rho$ with respect to $\mathcal{G}$, with elements $a_b \equiv \bra{\Phi_b}\rho\ket{\Phi_b}$ for any $b\in\mathbb{F}_2^n$. 
We note that the elements of $\mathbf{a}$ add up to unity due to the normalization constraint. 

\subsection{Predicting the graph-diagonal elements}

We begin by discussing the exact representation of the vector $\mathbf{a}$ of $\rho$. 
Connecting both the Bell state measurement results from the Bell sampling subroutine and the vector $\mathbf{a}$ are the estimates of the expectation values of the Pauli stabilizer elements. 
The following lemma expands $a_b$ in terms of these expectation values.

\begin{lemma}\label{lem:exactrep}
    Elements of $\mathbf{a}$ can be expressed as
    \begin{equation}\label{eq:a=linear(Tr)}
        a_b = \frac{1}{2^n} \left(\sum_{P\in \mathcal{S}_G^+} (-1)^{f_b(P)} \Tr(\rho P) \right).
    \end{equation}
    Here, the function $f_b\colon \mathcal{S}_G^+ \to \mathbb{F}_2$ is defined as
    \begin{equation}
        f_{b}(P) \equiv 
        \begin{cases}
            0 & \text{if $[\,P,U_b\,]=0$},  \\
            1 & \text{if $\{P,U_b\}=0$},
        \end{cases}
    \end{equation}
    where $U_b \equiv \bigotimes_{i=1}^{n} Z^{b_i}$ and $b \equiv \overline{b_1 \cdots b_{1<i<n} \cdots b_{n}}\in \mathbb{F}_2^{n}$ is an $n$-bit string.
\end{lemma}
We provide a detailed proof for this lemma in Appendix~\ref{app:Lemma-diagonal}. 
In essence, $a_b$ is a sum of the expectation values of the elements of the set $\mathcal{S}_G^+$, each term multiplied by a phase described by the unitary operator $U_b$ that generates $\ket{\Phi_b}$ (that is, due to phase flips incurred in $\ket{\Phi_G}$). 
We also note the case $b=0^n$ where the phase is unity for every stabilizer element $P$, which is accurate since the sum calculates the fidelity $F(\Phi_{G}, \rho) = \mel{\Phi_G}{\rho}{\Phi_G}$. 

Although the magnitudes of the expectation values of the stabilizer elements can be obtained from the Bell sampling protocol, there is still the problem of evaluating their signs. 
To address this, we employ the following Theorem to predict the signs deterministically. 
\begin{theorem}\label{thm:signs}
    If $ a_b > {1}/{2}$ for some $b$, then $\mathrm{sgn}(\Tr(\rho P)) = (-1)^{f_{b}(P)}$ for all $P\in \mathcal{S}_G^+$. In particular, if $a_{0^n}> 1/2$, then $\mathrm{sgn}(\Tr(\rho P)) = +1$ for all $P\in \mathcal{S}_G^+$.
\end{theorem}
\noindent We direct interested readers to Appendix~\ref{app:theorem-signs} for the details of the proof. 
This theorem follows from inverting Eq.~\eqref{eq:a=linear(Tr)} to obtain an exact representation of the Pauli expectation value $\Tr(\rho P)$ as a sum of $(-1)^{f_b(P)}a_b$ for all $b\in \mathbb{F}_2^n$. 
By imposing the condition $a_{b^*}>1/2$ for a particular $b^*$, the term $(-1)^{f_{b^*}(P)}a_{b^*}$ dominates the entire sum, which allows us to estimate the sign of $\Tr(\rho P)$ deterministically. 
In our specific scenario in which we compare the produced state $\rho$ with the desired state $\ket{\Phi_G}$, we can expect that the underlying noise is tolerable enough for a target network task so that the assumption $a_{0^n}> 1/2$ holds and that any state produced with $a_{0^n} \leq 1/2$ is useless for any practical use.
Under these conditions, the expectation values $\Tr(\rho P)$ can be obtained directly from the Bell sampling protocol, since all signs are positive following Theorem~\ref{thm:signs}.

\subsection{Sample complexity}

We now analyze the performance of BSQN under the assumption that the distributed state $\rho$ has fidelity to the target graph state $\ket{\Phi_G}$ greater than $1/2$. 
We focus on the $\ell_2$-norm of the $2^n$-length error vector $\Delta\mathbf{a} \equiv \mathbf{\hat{a}} - \mathbf{a}$, where $\mathbf{\hat{a}}$ is a vector of the estimations $\hat{a}_b$ of $a_b$. 
The following theorem presents the sample complexity required to bound this $\ell_2$-norm.

\begin{theorem}\label{thm:main}
    With probability no less than $1-\delta$, the vector $\mathbf{a}$ corresponding to the graph-diagonal elements of $\rho$ can be learned such that $\Vert \Delta\mathbf{a} \Vert_2 \leq \varepsilon$ using $\mathcal{O}\!\left((n + \log(1/\delta))/{\varepsilon^4}\right)$ samples.
\end{theorem}
\noindent Here, the notation $\Vert \cdot \Vert_p$ denotes the $\ell_p$-norm of a vector $\mathbf{v}$, with $\Vert \mathbf{v} \Vert_p \equiv \left(\sum_i \abs{v_i}^p\right)^{1/p}$.
We provide a detailed error analysis in Appendix~\ref{app:theorem-sample-complexity}. 
The $1/\varepsilon^4$ contribution is inherent to the Bell sampling subroutine, which is also consistent with previously known results such as in Refs.~\cite{huang2021information,chen2024optimal}. 
Notice that the $\ell_2$-norm of the difference $\Delta\mathbf{a}$ is proportional to the Hellinger distance \cite{canonne2020short} between the empirical and the true discrete distributions of the squared expectation of the stabilizer elements. 
We reiterate that this empirical discrete distribution is directly accessible through Bell sampling.

We now remark on the difference between our proposed protocol and the protocol in Ref.~\cite{huang2021information} to estimate this $\ell_2$-norm. 
The signs are fully estimated by majority voting in Ref.~\cite{huang2021information}, in which $2^n$ two-outcome observables are sequentially measured on $\mathcal{O}((n+\log(1/\delta))/\varepsilon^2)$ samples over a quantum memory. 
In total, $\mathcal{O}((n+\log(1/\delta))/\varepsilon^4)$ samples are consumed if we also take into account the contribution from Bell sampling. 
On the other hand, BSQN can determine the signs purely by using classical memory. 
 
\subsection{Measurements and classical post-processing}

From the previous subsection, we know that the number of measurements required by BSQN to learn $\mathbf{a}$ is, at worst, of the order $\mathcal{O}(n)$. 
We compare our result to a direct graph-diagonal estimation or DGE protocol. 
With DGE, all $2^n$ stabilizer elements are measured directly using individual copies of the graph state. The measurement results are then used to evaluate the diagonal entries through Eq.~\eqref{eq:a=linear(Tr)}. 
Unlike BSQN, the expectation value $\Tr(\rho P)$, both sign and magnitude, can be fully determined from the measurement results of DGE.
We note that the measurement complexity for DGE is in the order $\mathcal{O}(2^n)$. 
We can be smarter with the measurements that we perform depending on the graph structure. 
For example, we can combine measurements for different stabilizer elements, but we predict that such a strategy will not change the $n$-scaling. 
As such, there is a clear advantage in predicting diagonal elements when using BSQN.

We also remark on the classical post-processing stage of BSQN. Lemma~\ref{lem:exactrep} shows that each graph-diagonal element can be exactly expressed in terms of $2^n$ stabilizer measurements. 
This presents a strain in the classical post-processing stage, since the required memory and computational effort scale unfavorably with $n$. 
For example, if we need to distribute a graph state over $n=100$ nodes, naively calculating one diagonal element requires a classical memory that must accommodate about $2^{100}\simeq 10^{30}$ bits of information.
Taking into account all diagonal elements, we perform a linear transformation from the measurement results to the estimates $\hat{a}_b$ through Eq.~\eqref{eq:a=linear(Tr)}. 
This transformation, which is a classical Walsh-Hadamard transformation, will take $\mathcal{O}(n\, 2^n)$ steps. 
This scaling is exponential in $n$ and is therefore a major bottleneck in computation.

\subsection{Estimation with few stabilizer elements}

We see in the previous subsection the potential challenges in resource overhead and in classical post-processing when using BSQN for state characterization. 
These challenges are even more pronounced for graph states distributed over a much larger network.

A practical approach to alleviate this scalability problem is to extract enough information about a subset of diagonal elements (and, in particular, the fidelity with respect to $\ket{\Phi_G}$) to provide a meaningful characterization. 
However, notice that following Theorem~\ref{thm:main}, learning the diagonal elements individually still retains an $n$-dependence on sample complexity since we have the inequalities $\abs{\Delta a_b}\leq \Vert \Delta \mathbf{a} \Vert_{\infty} \leq \Vert \Delta \mathbf{a} \Vert_2$, where ${\Delta a_b} = \hat{a}_b - a_b$. 

Here, we show that we can estimate a diagonal entry in a more efficient way. 
Inspired by the direct fidelity estimation approach in Ref.~\cite{flammia2011direct}, we can effectively remove this dependency in $n$ by employing random sampling of the stabilizer elements considered during the estimation. 
\begin{theorem}\label{thm:fidelity_random}
    Suppose that a learner has access to a classical memory of size $M=\mathcal{O}(\log(1/\delta)/\varepsilon^2)$, 
    independent of the graph order,
    which limits the number of stabilizer measurement settings that can be stored and processed.
    With probability no less than $1-2\delta$, an element of $\mathbf{a}$ can be learned with error $2\varepsilon$ using $\mathcal{O}(\log(1/\varepsilon^2\delta)/\varepsilon^4)$ samples.
\end{theorem}
\noindent We note that the measurement settings are selected randomly and uniformly and that these settings can be repeated. 
The details of the proof are outlined in Appendix~\ref{app:random_sampling}. 
The proof defines an unbiased estimator of $a_b$, which is written as a sum of $M$ independent infinite-precision estimates $(-1)^{f_{b\oplus b^{\ast}}(P)} \abs{\Tr(\rho P)}$ for a single coefficient $a_{b^{\ast}}>1/2$. 
This unbiased estimator is then approximated using the measurement results from the Bell sampling subroutine. 
Using this characterization method, we measure each randomly chosen stabilizer element with $\mathcal{O}(\log(1/\varepsilon^2\delta)/M\varepsilon^4)$ samples of $\rho$.

\section{Pauli error detection}
\label{sec:errordetection}

We note that BSQN characterizing only the graph-diagonal elements already provides a large amount of meaningful information about the distributed state. 
To see this, suppose that $\mathcal{N}$ is a noisy $n$-partite Pauli channel.
With the task of distributing the graph state $\ket{\Phi_G}$, we have the noisy state
\begin{equation}
    \mathcal{N}(\Phi_G) = \sum_{P\in \mathcal{P}_n^+}\lambda_P P\ketbra{\Phi_G}{\Phi_G}P,
\end{equation}
where the parameters $\lambda_P$ are the Pauli error rates.
This state can be written on the graph state basis as follows:
\begin{equation}
    \mathcal{N}(\Phi_G)=\sum_{b\in \mathbb{F}_2^n}\left[\sum_{P\in (U_b\mathcal{S}_G)^+}\lambda_P\right]\ketbra{\Phi_b}{\Phi_b}.
\label{eq:pauli-error-in-graph-basis}
\end{equation}
Here, $(U_b \mathcal{S}_G)^+$ represents the 
coset $U_b \mathcal{S}_G$ modulo the overall phase.
The elements of $U_b \mathcal{S}_G$ 
map $\ket{\Phi_{0^n}}$ to another element $\ket{\Phi_b}$ of $\mathcal{G}$, up to a global phase.
We can then reinterpret $\mathcal{N}(\Phi_G)$ as an ensemble of graph states $\ket{\Phi_b}$ belonging to $\mathcal{G}$, with each $a_b$ the associated ensemble probability equal to the $\lambda_P$-sum within the square brackets. 
Note that each element of $\mathcal{G}$ is described by $U_b$, which is a representative element of $U_b \mathcal{S}_G$. 
This $U_b$ describes a specific set of only dephasing errors incurred in the graph state.
Moreover, all the $2^n$ elements of $U_b \mathcal{S}_G$ transform $\ket{\Phi_G}$ into $\ket{\Phi_b}$. 
The diagonal element $a_b$ then corresponds to the probability of error in which all elements of $(U_b \mathcal{S}_G)^+$ act on $\ket{\Phi_G}$.
This means that our approach is applicable to arbitrary Pauli channels, not just dephasing errors.
However, we point out that BSQN cannot disambiguate the contribution of each element in $(U_b \mathcal{S}_G)^+$, since $a_b$ is described as a sum of different Pauli error rates.

\begin{figure}[t]
    \centering
    \includegraphics[width=1.0\linewidth]{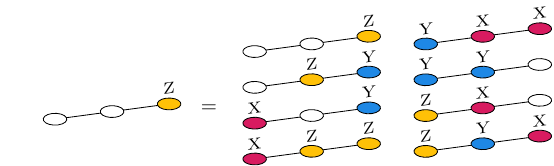}
    \caption{\label{fig:noise-equivalence}
    Effects of general Pauli channels acting on graph states can be captured by dephasing errors only. For example, the effect of a single $Z_3$ error acting on a linear graph state is equivalent to eight distinct Pauli errors.}
\end{figure}

To illustrate this point with an example, we consider a $3$-qubit-linear graph state $\ket{\Phi_G}$ with the stabilizer $\mathcal{S}_G = \langle X_1Z_2, Z_1X_2Z_3, Z_2X_3\rangle$.
The dephasing error $U_b=Z_3$ on the third qubit transforms the graph state into $Z_3\ket{\Phi_G}$. 
However, $Z_3$ is not the only operation that yields this state.
The same state can be obtained by applying $Z_1X_2$ to $\ket{\Phi_G}$.
In fact, there are six other distinct Pauli errors that we list in Fig.~\ref{fig:noise-equivalence} that produce the same state, which are all elements of $U_b\mathcal{S}_G = Z_3\mathcal{S}_G$ (modulo signs) in the $n$-Pauli group.

On the other hand, we can conveniently ignore the nonzero off-diagonal elements caused by coherent (non-Pauli) noise: we can always depolarize any noisy graph state into an ensemble of graph states by a sequence of (device) local operations and classical communication, with diagonal elements preserved as ensemble probabilities \cite{Aschauer2005Multiparticle}. 
These ensembles of graph states have been well studied, for example, in the context of multipartite entanglement distillation \cite{Dur2003Multiparticle,Kruszynska2006entanglement}.

\section{Numerical Results}\label{sec:results}

\begin{figure*}
    \centering
    \includegraphics[width=\linewidth]{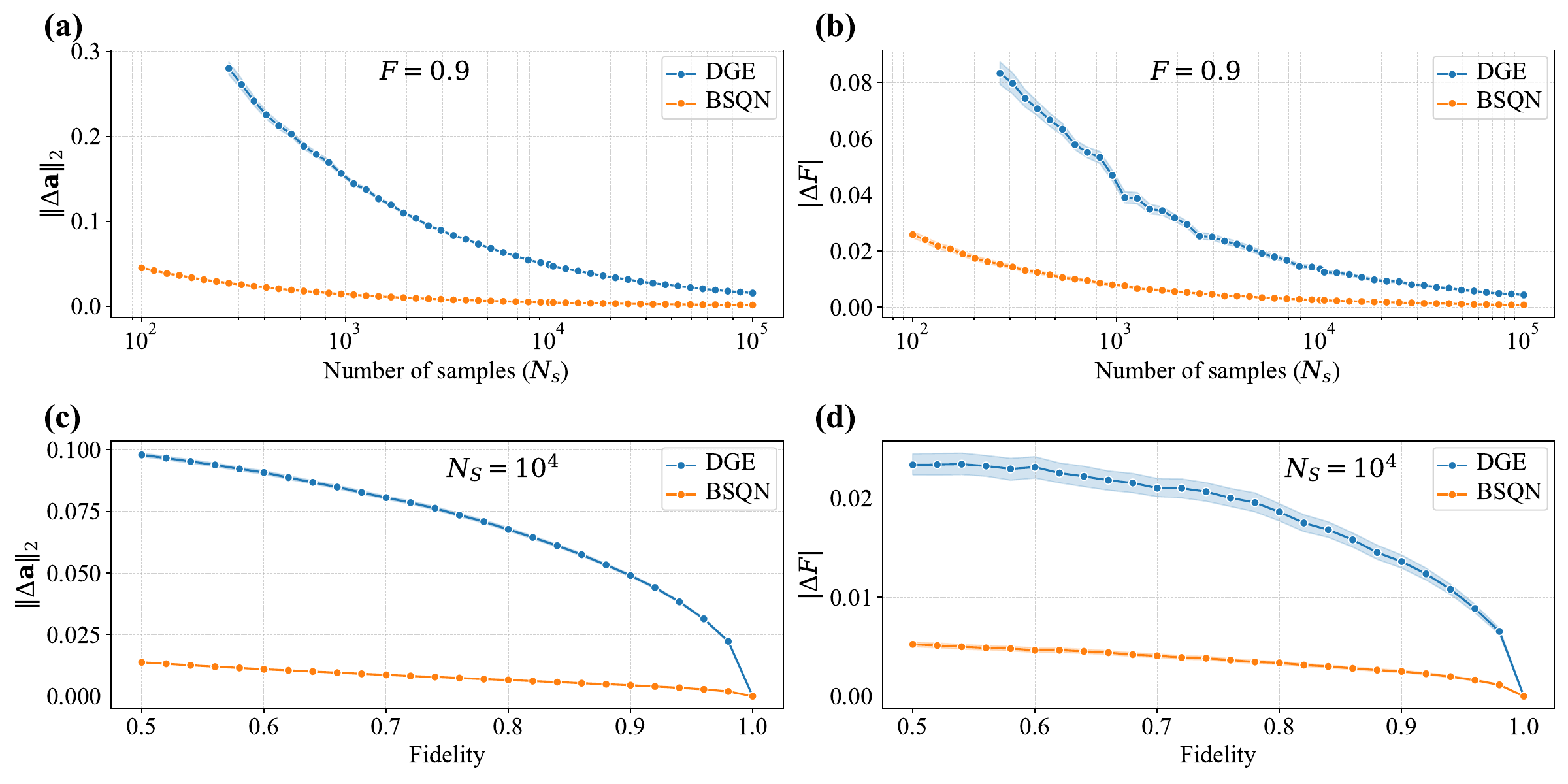}
    \caption{Performance comparison between BSQN and DGE for an $8$-node complete graph state. The top row (a, b) shows the $2$-norm error $\Vert \Delta \mathbf{a} \Vert_2$ and fidelity error $\abs{\Delta F}$ as a function of the number of copies $N_s$, for a fixed fidelity of $F=0.9$. The bottom row (c, d) shows the same errors as a function of fidelity $F$, for a fixed $N_s=10^4$. The markers show the mean while the envelopes (not visible in (c)) show the standard deviation.}
    \label{fig:norm_dF}
\end{figure*}

In this section, we present numerical results that demonstrate the performance and scalability of BSQN compared to DGE.
We analyze the effectiveness of the protocol to learn the whole diagonal vector $\mathbf{a}$ of $\rho$, as well as the specific task of estimating the fidelity.
(The details of the implementation of DGE and the simulation are provided in Appendix~\ref{app:dge_description} and~\ref{app:simulation_procedure}, respectively).

\subsection{Simulation setup}

We begin by describing the noise models that we picked for our investigation.
Throughout this discussion, we denote $a_{0} = F$ for the fidelity with $\ket{\Phi_G}$.
We consider an $n$-partite Pauli noise model $\mathcal{N}_{F,\mathbf{r}}$ parameterized by a stochastic vector $\mathbf{r} \in \mathbb{R}^{2^n -1}$.
The corresponding distributed state is
\begin{equation}
    \mathcal{N}_{F,\mathbf{r}}(\Phi_{G}) = F\ketbra{\Phi_{G}}{\Phi_{G}} +(1-F)\!\!\sum_{b\in \mathbb{F}_{2}^n \setminus 0^n } r_{b} \ketbra{\Phi_b}{\Phi_b}.
\end{equation}
For the first noise model, which we label $\mathcal{N}_1$ and refer to as the \emph{depolarizing noise channel}, we consider a uniform noise model where the error is completely unbiased.
In this case, $r_b = 1/(2^{n}-1)$ for all $b\neq 0$, which means that the state is a uniform mixture of all other graph basis elements.
We describe the second model $\mathcal{N}_2$, which we call the \emph{single-qubit dephasing channel}, such that the $n$ qubits are under $n$ i.i.d. dephasing channels (i.e., each qubit has its own identical noise channel) with a dephasing parameter $\mu$.
Then, the fidelity is $F = (1-\mu)^n$, and the stochastic vector elements are $r_b = {\mu^{h}(1-\mu)^{n-h}}/{(1-(1-\mu)^n)}$, where $h=\mathrm{wt}(b)$ is the Hamming weight of $b$.
Finally, we consider a maximally biased noise model $\mathcal{N}_3$, which we refer to as \emph{bimodal noise}, where $r_b=1$ for a single $b\neq 0$, and zero otherwise.

In all simulation settings, we consider graph states defined by an $n$-qubit complete graph. 
We selected this topology as it represents a standard benchmark for multipartite entanglement, corresponding to a state locally equivalent to an $n$-qubit GHZ state. 
Since the BSQN estimation framework is independent of the underlying graph topology (see Appendix~\ref{app:graph-independence} for details), the performance and scalability results presented for the complete graph are representative and can be generalized to states with other structures, such as path or star graphs.
The complete graph structure also allows us to devise a strategy to combine multiple stabilizer elements into a single measurement setting making DGE more sample efficient (see Appendix~\ref{app:simulation_procedure} for details).

\subsection{Errors of estimation}

We explicitly study the estimation errors associated with the diagonal vector, $\Vert \Delta \mathbf{a} \Vert_p$, and the fidelity, $\abs{\Delta F}$, of $\rho$ as our figures of merit.
Here, we denote by $\Delta F = \hat{F}-F$ the fidelity error, where $\hat{F}$ is the estimate for $F$.

Figure~\ref{fig:norm_dF} shows the numerical analysis for the complete graph with $n=8$.
In Figures~\ref{fig:norm_dF}(a) and \ref{fig:norm_dF}(b), we numerically estimate vector errors in terms of the $2$-norm and the fidelity errors against the total number of copies $N_s$ consumed.
We vary $N_s$ from $10^2$ to $10^5$ for BSQN.
For DGE, we vary $N_s$ from a minimum of $268$ (corresponding to approximately two shots per stabilizer element) up to $10^5$.
We initialize $\rho$ with a fidelity of $0.9$ under the $\mathcal{N}_1$ depolarizing noise model and perform $1000$ trials.
The results indicate that BSQN estimates the distributed states with high accuracy even at low sample counts---a regime DGE cannot enter due to its higher minimum sample requirement.
Given these experimental parameters, we observe that DGE requires significantly more samples to achieve comparable confidence levels for both $\mathbf{a}$ and fidelity, demonstrating the significant advantage of BSQN for estimating the full graph-diagonal vector.

In Figures~\ref{fig:norm_dF}(c) and \ref{fig:norm_dF}(d), we analyze the errors as a function of fidelity, given a fixed number of copies ($N_s=10^4$).
We repeat the implementation for $1000$ trials.
The results again demonstrate that our protocol is sample efficient over DGE, especially when estimating the diagonal elements of the full vector $\mathbf{a}$ as the states become noisier (i.e., as the fidelity decreases).
Our protocol also shows significantly better performance for estimating fidelities across the entire fidelity range.

We note that DGE can be implemented in multiple ways, as there are various ways to perform single-qubit measurements on a state that allow us to extract multiple expectation values of stabilizer elements via a single measurement setting.
For the DGE results shown in Figure~\ref{fig:norm_dF}, we chose an implementation that maximizes the number of shots allocated equally to each stabilizer element within the budget $N_s$.
A different implementation of DGE will give different results (see Appendix~\ref{app:errors-with-different-dge} for details), where one can estimate the fidelity better while sacrificing the accuracy for the full graph-diagonal vector, resulting in an increased 2-norm error.

\subsection{Scalability with the number of qubits}

\begin{figure}[t]
    \centering
    \includegraphics[width=\columnwidth]{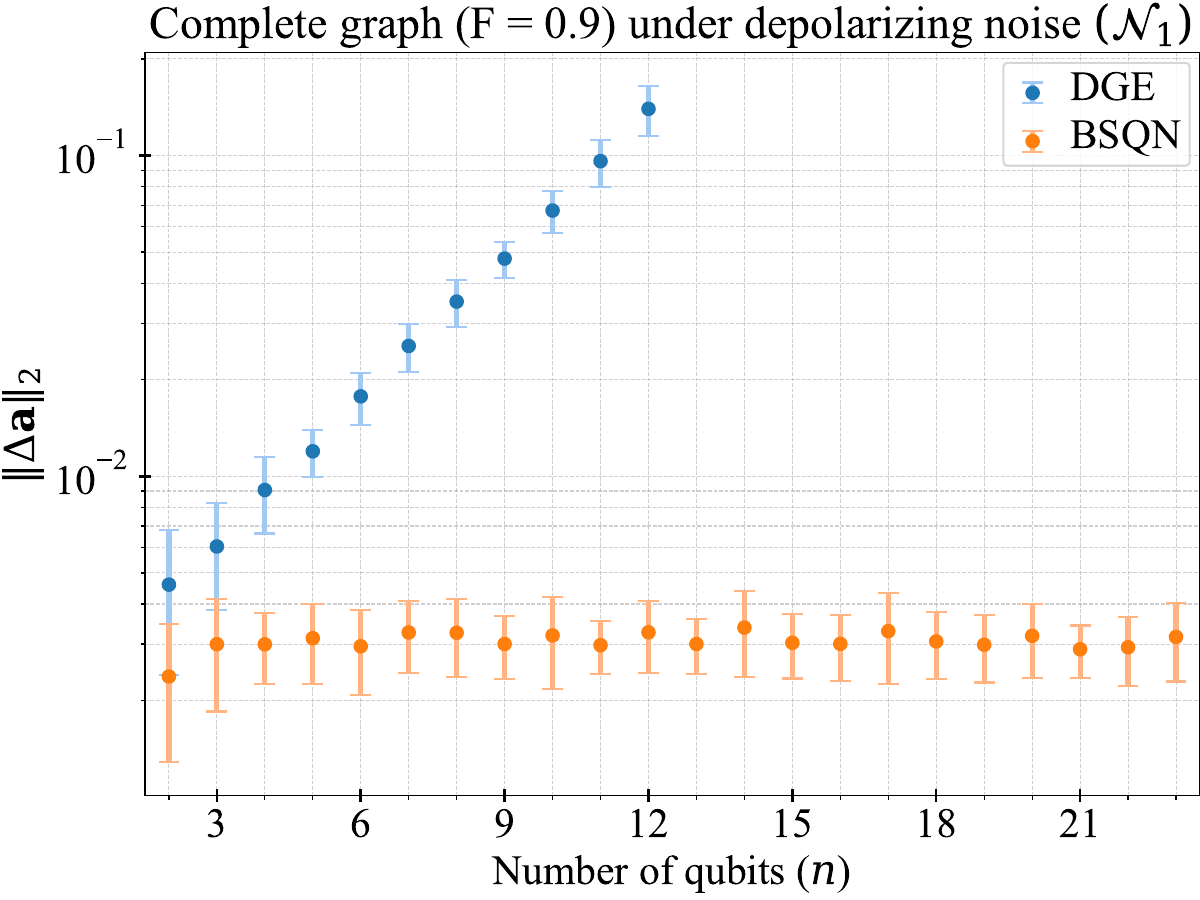}
    \caption{Scalability comparison of BSQN and DGE with fixed number of copies of the graph state, $N_s = 2 \times 10^4$. The plot shows the $2$-norm estimation error $\Vert \Delta \mathbf{a} \Vert_2$ as a function of the number of qubits $n$. The states $\rho$ are noisy $n$-node complete graph states with fidelity $F = 0.9$ under depolarizing noise $\mathcal{N}_1$.}
    \label{fig:N_complexity}
\end{figure}

So far, we have analyzed the performance of BSQN in terms of accuracy of estimation given modest resource consumption.
We now numerically assess the scalability of our proposed protocol with respect to the order of the graph, where we once again use the $\ell_p$ norms as figures of merit.
We compare the numerical results with the pessimistic scaling in Theorem~\ref{thm:main}.

Figure~\ref{fig:N_complexity} shows the $2$-norm $\Vert \Delta \mathbf{a} \Vert_2$ as a function of the number of qubits $n$ in a complete graph.
In this figure, we choose $\mathcal{N}_1$ as our noise model.
We fix the fidelity at $F = 0.9$ and repeat the estimation protocol for $25$ trials with $N_s = 20\thinspace000$.
We compare this with the DGE defined earlier.
The results already demonstrate the exponential advantage of BSQN in scalability (see the vertical axis on logarithmic scale).
We note that for a fixed number of copies consumed, Theorem~\ref{thm:main} predicts that the estimation error behaves as $\mathcal{O}(n^{1/4})$, which grows slowly on the logarithmic scale.
Interestingly, we observe that the $2$-norm error appears constant for relatively small values of $n$, which could suggest an instance of $n$-independence.

\subsection{Performance of estimation with random sampling}

\begin{figure*}[th]
    \centering
    \includegraphics[width=1\linewidth]{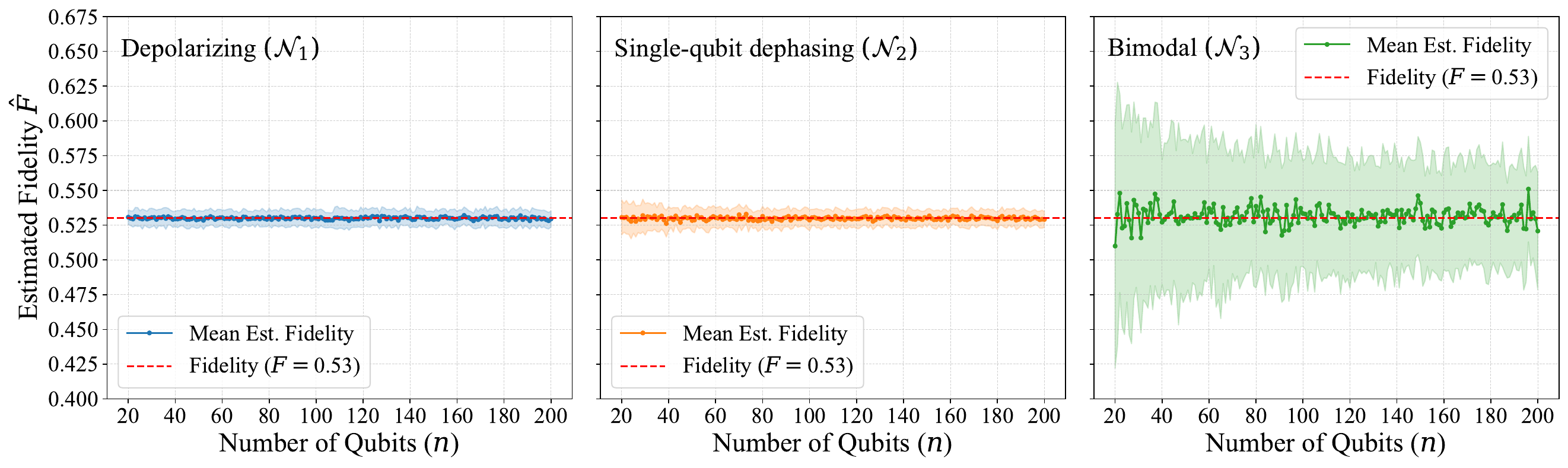}
    \caption{Scalability of the random sampling fidelity estimation protocol (Theorem~\ref{thm:fidelity_random}). The plots show the mean estimated fidelity (solid line) and standard deviation (envelope) as a function of the number of qubits $n$ with the number of sampled stabilizer elements $M = 2n$ and fixed fidelity $F = 0.53$, under different noise models.}
    \label{fig:fidelity_random_increasing_qubits}
\end{figure*}

\begin{figure*}[th]
    \centering
    \includegraphics[width=\linewidth]{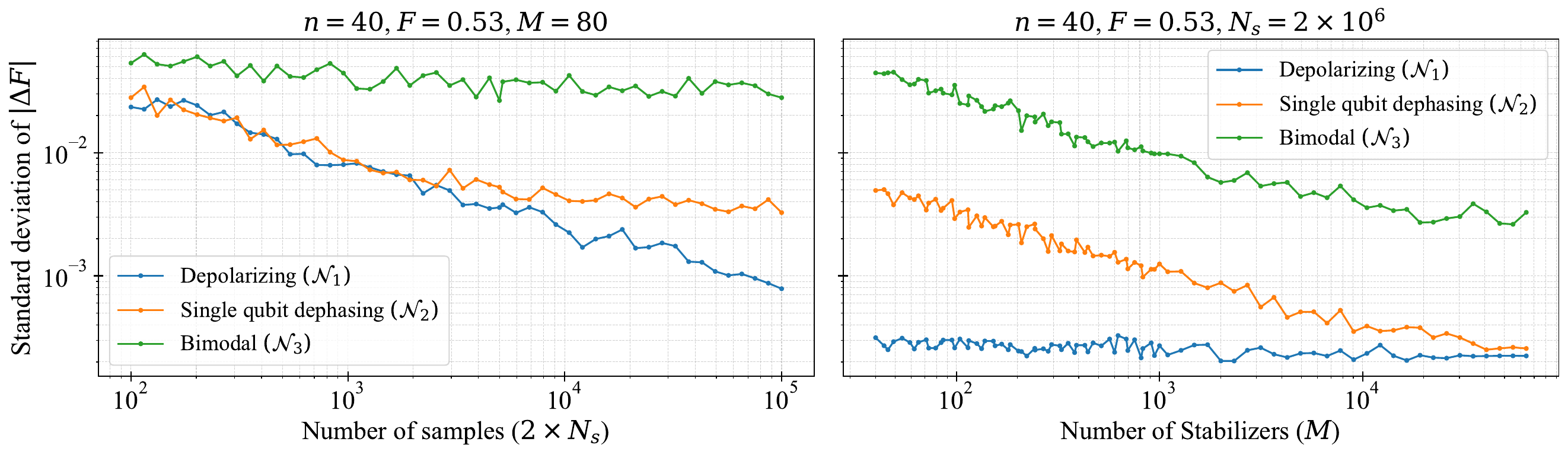}
    \caption{Estimator precision for the random sampling protocol (Theorem~\ref{thm:fidelity_random}) as a function of different resources. The plots show the standard deviation of the fidelity errors ($\vert \Delta F \vert$) while varying (left) the number of copies consumed $N_s$ and (right) the number of randomly sampled stabilizer elements $M$.}
    \label{fig:fidelity_random_increasing_shots_and_stabs}
\end{figure*}

We now discuss the benchmarking results of the random sampling protocol (Theorem~\ref{thm:fidelity_random}), which is designed to efficiently estimate individual diagonal elements, such as fidelity (corresponding to $a_0$).

First, we investigate the protocol's scalability with the system size.
Figure~\ref{fig:fidelity_random_increasing_qubits} illustrates the performance of the fidelity estimation protocol from Theorem~\ref{thm:fidelity_random} on complete graph states of varying orders, from $n=20$ to $n=200$.
For these simulations, we set the initial fidelity to 0.53 (chosen to represent a more difficult estimation regime as seen in Figure~\ref{fig:norm_dF}), use a fixed number of copies of 10\thinspace000 per trial, and sample $M=2n$ stabilizer elements.
In each subplot, the solid line indicates the mean of the fidelity estimates, while the shaded envelope represents the standard deviation.
The results show that the bimodal noise model ($\mathcal{N}_3$) yields the largest standard deviation, which is approximately 0.07 for smaller graphs and decreases to roughly 0.045 as the graph order increases into the $n=100$ to $n=200$ range.

In Figure~\ref{fig:fidelity_random_increasing_qubits}, we established that the mean of the fidelity estimates $\hat{F}$ is accurate (i.e., it approaches the true fidelity $F$).
We now turn our analysis to the estimator's precision and its associated resource requirements.
We fix the system size at $n=40$ with an initial fidelity of $F=0.53$ and study the standard deviation of the error of the estimates, as shown in Figure~\ref{fig:fidelity_random_increasing_shots_and_stabs}.
The left subplot shows the effect of increasing the number of copies from $2 \times 10^2$ to $2 \times 10^5$ while keeping the number of stabilizer elements fixed at $M=80$ ($M = 2n$).
For the depolarizing model ($\mathcal{N}_1$), the standard deviation decreases significantly, and a similar, though less pronounced, trend is observed for the single-qubit dephasing noise model ($\mathcal{N}_2$).
In contrast, the estimator's precision for the bimodal noise model ($\mathcal{N}_3$) shows slight improvement within this range of number of copies (noted that the vertical axis is in logarithmic scale).
The right subplot illustrates the impact of increasing the number of sampled stabilizer elements, $M$, from 40 to $40^3$ ($\approx 6.4 \times 10^4$), with a fixed number of copies, $N_s=2 \times 10^6$.
We emphasize that this range is an extremely small subset of the total $2^{40} \approx 10^{12}$ stabilizer elements.
Here, the precision for $\mathcal{N}_1$ is largely insensitive to an increase in $M$, while for $\mathcal{N}_2$ and $\mathcal{N}_3$, we observe a slow decreasing trend in the standard deviation.
For $\mathcal{N}_2$, the standard deviation also seems to plateau after sampling $M \approx 8 \times 10^4$ stabilizer elements (still a small fraction of the total number of elements) similar to the behavior observed for $\mathcal{N}_1$, and both error floors appear to almost converge.

Overall, these results indicate that different noise models have different resource requirements.
The depolarizing model ($\mathcal{N}_1$) is primarily sensitive to the number of copies, while the single-qubit dephasing model ($\mathcal{N}_2$) benefits from an increase in both copies and stabilizers.
The bimodal noise ($\mathcal{N}_3$) appears most difficult to learn the fidelity of the state, as increasing only the number of copies or only the number of stabilizers provides limited improvement.

We conjecture that estimating the fidelity requires sampling a sufficient number of stabilizer elements, $M$, to adequately characterize the noise distribution.
The results for $\mathcal{N}_1$ (right subplot, $N_s = 2 \times 10^5$) suggest that this requirement is quickly saturated for simple noise models; beyond this point, increasing $M$ does not provide a significant benefit, and the precision is limited only by the number of copies.
The single-qubit dephasing model ($\mathcal{N}_2$) represents an intermediate case, where the precision improves with $M$ before plateauing, suggesting a larger but still achievable saturation threshold.
Conversely, for ``hard-to-learn'' distributions like the maximally biased noise of $\mathcal{N}_3$, a much larger $M$ is likely required, and we do not observe saturation within our tested range.
Only after this threshold is met would we expect to see a significant decrease in the standard deviation by increasing the number of copies (similar to the trend seen for $\mathcal{N}_1$ in the left subplot).
Such cases may require a simultaneous increase in both resources (number of copies of the state and the time for post-processing the measurement results), possibly approaching the pessimistic bounds of our proof.
However, for simpler models, these experiments show that the practical resource costs can be quite modest.
We leave a more exhaustive numerical study of these resource trade-offs for future work.

\section{Summary and outlook}\label{sec:summary}

In this work, we proposed Bell Sampling for Quantum Networks (BSQN) to efficiently learn the graph-diagonal elements of noisy graph states distributed across a network. 
This protocol employs Bell sampling as a subroutine that measures stabilizer expectation values using two simultaneous copies of the state. 
We have analytically shown that all the diagonal elements in the graph basis can be accurately estimated solely from classically post-processing the measurement statistics obtained from Bell sampling provided that the fidelity is greater than a half. 
We note the experimental demonstrations that produced states that are local unitarily equivalent to graph states that already pass this one-half fidelity threshold, although with low count rates \cite{Lu2007experimental,Tokunaga08,Yao2012,Wang16,Wang18,Gong19,Thomas2022efficient}. 
Typically and under current experimental capabilities, certifying these produced states is done using fidelity-based entanglement witnesses \cite{Guhne2009entanglement}. 
Entanglement witnesses in principle provide less information 
than our BSQN protocol. 
As technology improves (e.g., having the capability of simultaneously measuring two copies), we see that BSQN will be a more suitable alternative for certification.

We also gave explicit bounds for the sample complexity of the protocol. 
In particular, we found that the number of resources required to simultaneously learn all diagonal elements scales linearly with the order of the graph (that is, the size of the network).
Moreover, by only measuring a constant number of randomly selected stabilizer expectation values, we have shown that a single diagonal element can be learned using our protocol with a sample complexity independent of network size. 
We observed that both strategies are sample-efficient over the resource intensive direct graph-diagonal estimation that directly measures the stabilizer elements, with the efficiency more pronounced as the order of the graph increases. 
However, numerical analysis demonstrates that BSQN can perform even with a modest resource overhead for some simple noise models.
In particular, we find evidence of much better complexity scaling for these noise models than was promised following our analytical error bounds.

Although BSQN only performs partial characterization, since it only outputs the estimates for the graph-diagonal elements of the distributed state, we have argued that this characterization can already provide ample information on all the possible Pauli errors incurred during distribution. 
We also point out the relatively simple implementation of the protocol on the level of the local devices, since we only perform Bell state measurements at every node.
These showcase the potential of our proposed protocol as a sample- and implementation-efficient error detection tool for near-term networks.  

\section{Acknowledgments}

The authors thank William John Munro, Nana Liu, Tim Coopmans, Kenneth Goodenough, Vedran Dunjko, Lorenzo La Corte, and Xiaoyu Liu for useful discussions during the project.
The project was supported by the JST Moonshot R\&D
program under Grant No.~JPMJMS226C.
Z.J.W. is supported by the ERC Consolidator Grant BeMAIQuantum (Grant No.~101124342) and by the Dutch National Growth Fund (NGF) as part of the Quantum Delta NL program. 
J.T. further acknowledges support from the European Union’s Horizon Europe research and innovation programme via the ERC Starting Grant FINE-TEA-SQUAD (Grant No.~101040729).

The views and opinions expressed here are solely those of the authors and do not necessarily reflect those of the funding institutions. 
Neither of the funding institutions can be held responsible for them.

\vspace{1.5cm}

\onecolumngrid

\appendix

\section*{Appendix}

\section{Proof of Lemma \ref{lem:exactrep}}\label{app:Lemma-diagonal}

In this section, we give the proof of Lemma \ref{lem:exactrep}, which we recall for the reader:

\renewcommand{\thelemma}{\ref{lem:exactrep}}
\begin{lemma}
    Elements of $\mathbf{a}$ can be expressed as
    \begin{equation}
        a_b = \frac{1}{2^n} \left(\sum_{P\in \mathcal{S}_G^+} (-1)^{f_b(P)} \Tr(\rho P) \right).
    \end{equation}
    Here, the function $f_b\colon \mathcal{S}_G^+ \to \mathbb{F}_2$ is defined as
    \begin{equation}
        f_{b}(P) \equiv 
        \begin{cases}
            0 & \text{if $[P,U_b]=0$},  \\
            1 & \text{if $\{P,U_b\}=0$},
        \end{cases}
    \end{equation}
    where $U_b \equiv \bigotimes_{i=1}^{n} Z^{b_i}$.
\end{lemma}
\renewcommand{\thelemma}{\arabic{lemma}} 

\begin{proof}
    For concreteness, let us define the unitary (Clifford) operator $U_G \equiv \prod_{(u,v)\in E} CZ_{u,v}$ associated with graph $G=(V,E)$, 
    and the Pauli operator $U_b \equiv\bigotimes_{i=1}^{n} Z^{b_i}$ where $n = \abs{V}$ is the number of vertices of the graph. 
    We define $b$ as a bitstring of length $n$, i.e., $b \equiv \overline{b_1 \dots b_n} \in \{0,1\}^n$. 
    Let us also rewrite the graph state $\ket{\Phi_G}$ associated with $G$ as $\ket{\Phi_{0^n}}$, with ${0^n}$ denoting the bitstring $\overline{0\dots 0}$. 
    That is,
    \begin{equation}
        \ket{\Phi_{0^n}} \equiv \ket{\Phi_G} = U_G \ket{+}^{\otimes V}.
    \end{equation}
    The other elements of the graph-state basis $\mathcal{G}$, where $\abs{\mathcal{G}}=2^{n}$, can also be obtained with the following: 
    \begin{equation}
        \ket{\Phi_{b}} = U_G U_b \ket{+}^{\otimes V} = U_b \ket{\Phi_{0^n}}\in \mathcal{G}.
    \end{equation}
    Here, we have $U_b=\bigotimes_{i=1}^{n}Z^{b_i}$, and we use the fact that $U_G$ commutes with $U_b$.

    We define the $n$-Pauli group modulo phases as $\mathcal{P}_n^+ = \{I,X,Y,Z\}^n$, with size $\abs{\mathcal{P}_n^+}=4^n$. We use the elements of $\mathcal{P}_n^+$ to parameterize any $n$-qubit density matrix $\rho$ into 
    \begin{equation}
        \rho = \frac{1}{2^n}\sum_{P\in \mathcal{P}_n^+}\Tr(\rho P)P.
    \end{equation}
    By definition, $\ket{\Phi_{0^n}}$ is stabilized by the group $\mathcal{S}_G = \langle S_1, S_2, \cdots, S_n \rangle$, where the generators $S_v$ are defined in Eq.~\eqref{eq:graph_state_stabilizer}.
    Here, we construct the set $\mathcal{S}_G^+$ by taking all elements of $\mathcal{S}_G$ and discarding their overall phases.
    Calculating for the diagonal elements $a_{b} \equiv \bra{\Phi_b} \rho \ket{\Phi_b}$ in $\mathcal{G}$, we find
    \begin{equation}
        a_b = \bra{\Phi_{b}} \frac{1}{2^n}\sum_{P\in \mathcal{P}_n^+}\Tr(\rho P)P \ket{\Phi_{b}} = \frac{1}{2^n}\sum_{P\in\mathcal{P}_n^+}\Tr(\rho P)\mel{\Phi_{0^n}}{U_b P U_b}{\Phi_{0^n}}.
    \end{equation}
    Since $U_b$ consists solely of Pauli-$Z$ operators, $U_b P U_b$ equals either $P$ or $-P$. 
    Importantly, $\langle \Phi_{0^n} | P | \Phi_{0^n} \rangle$ is nonzero only if $P \in \mathcal{S}_G^+$: if $P \notin \mathcal{S}_G$, then $P$ must belong to some (equivalence class) $(U_b\mathcal{S}_G)^+$ in $\mathcal{P}_n^+$ whose elements bring $\ket{\Phi_{0^n}}$ to another element $\ket{\Phi_b}$ (ignoring the overall phase) of $\mathcal{G}$, and therefore $\langle \Phi_{0^n} | P | \Phi_{0^n} \rangle = 0$. 
    Consequently, we obtain
    \begin{equation}
        a_b = \frac{1}{2^n}\sum_{P\in \mathcal{S}_G^+} (-1)^{f_b(P)} \Tr(\rho P),
    \end{equation}
    where we define the function $f_b\colon \mathcal{S}_G^+ \to \mathbb{F}_2$,
    \begin{equation}
        f_{b}(P) \equiv 
        \begin{cases}
            0 & \text{if $[\,P,U_b\,]=0$},  \\
            1 & \text{if $\{P,U_b\}=0$},
        \end{cases}
    \end{equation}
    for some bit string $b$ and some Pauli operator $P$.
\end{proof}
    
\section{Proof of Theorem~\ref{thm:signs}}\label{app:theorem-signs}

Let us now present the proof of Theorem~\ref{thm:signs}, which we restate for convenience.
\renewcommand{\thetheorem}{\ref{thm:signs}}
\begin{theorem}
    If $ a_b > {1}/{2}$ for some $b$, then $\mathrm{sgn}(\Tr(\rho P)) = (-1)^{f_{b}(P)}$ for all $P\in \mathcal{S}_G^+$. In particular, if $a_{0^n}> 1/2$, then $\mathrm{sgn}(\Tr(\rho P)) = +1$ for all $P\in \mathcal{S}_G^+$.
\end{theorem}
\renewcommand{\thetheorem}{\arabic{theorem}} 

\begin{proof}
    For this proof, we define the column vector $\mathbf{a}$ of length $2^n$ whose elements are the diagonal elements of $\rho$ in the basis $\mathcal{G}$. 
    We also define the column vector $\mathbf{w}$ of length $2^n$ whose elements are $\Tr(\rho P)$, where $P\in \mathcal{S}_G^+$. 
    We can then rewrite Eq.~\eqref{eq:a=linear(Tr)} into its matrix form $\mathbf{a} = \mathbf{Q} \mathbf{w}$, where
    \begin{align}
       \mathbf{a} &= (a_{\overline{0\cdots0}},\ \cdots,\ a_{b},\ \cdots,\ a_{\overline{1\cdots1}})^T,\ 
       b = \overline{b_1 \dots b_n} \in \{0,1\}^n \\
       \mathbf{w} &= (1,\ \Tr(\rho P_1),\ \cdots,\ \Tr(\rho P_i),\ \cdots,\ \Tr(\rho P_{2^n-1}))^T, \ P_i \in \mathcal{S}_G^+, \ P_0 \equiv I^{\otimes n}
    \end{align}
    \begin{equation}
       \mathbf{Q} \equiv \frac{1}{2^n}
       \begin{pmatrix}
            (-1)^{f_{\overline{0\cdots0}}(I^{\otimes n})} & (-1)^{f_{\overline{0\cdots0}}(P_1)}  & (-1)^{f_{\overline{0\cdots0}}(P_2)} & \cdots & (-1)^{f_{\overline{0\cdots0}}(P_{2^n-1})} \\
            (-1)^{f_{\overline{0\cdots1}}(I^{\otimes n})} & (-1)^{f_{\overline{0\cdots1}}(P_1)} & (-1)^{f_{\overline{0\cdots1}}(P_2)} & \cdots & (-1)^{f_{\overline{0\cdots1}}(P_{2^n-1})} \\
            \vdots & \vdots & \vdots & \ddots & \vdots\\
            (-1)^{f_{\overline{1\cdots1}}(I^{\otimes n})} & (-1)^{f_{\overline{1\cdots1}}(P_1)} & (-1)^{f_{\overline{1\cdots1}}(P_2)} & \cdots & (-1)^{f_{\overline{1\cdots1}}(P_{2^n-1})} \\
       \end{pmatrix}
       \equiv \frac{1}{2^n} \mathbf{Q_0}. \label{eq:Q}
    \end{equation}
    It should be noted that the inverse of $\mathbf{Q}$ exists, that is, $\mathbf{Q}^{-1} = \mathbf{Q_0}^{T}$. 
    To show this, consider the matrix $\mathbf{A} = \frac{1}{2^n} \mathbf{Q_0} \mathbf{Q_0}^{T}$, with matrix elements
    \begin{equation}
        A_{b,b'} = \frac{1}{2^n} \sum_{P\in \mathcal{S}_G^+} (-1)^{f_{b}(P) + f_{b'}(P)}=\frac{1}{2^n} \sum_{P\in \mathcal{S}_G^+} (-1)^{f_{b\oplus b'}(P)}.
    \end{equation}

    Suppose that $\mathcal{U}_{b\oplus b'} = Z_v \otimes \mathcal{Z}_{V-\{v\}}$ for some vertex $v$, where $\mathcal{Z}_{V-\{v\}}$ is some tensor product of $Z$'s and $I$'s over the vertex set ${V-\{v\}}$. 
    Then, $f_{b\oplus b'} = 0$ if $P\neq X_v \otimes P_{V-\{v\}}$ or $P\neq Y_v \otimes P_{V-\{v\}}$ (up to a phase), where $P_{V-\{v\}}$ is a tensor product of Pauli operators over ${V-\{v\}}$. 
    The total number of such $P$'s in $\mathcal{S}_G$ is $\sum_{k=1}^{n-1}\binom{n-1}{k} = 2^{n-1}$ (i.e., we ignore all elements that are generated by $X_v\otimes \bigotimes_{u\in \mathrm{NN}(v)}Z_u$). 
    Also, we have $f_{b\oplus b'} = 1$ for $\abs{\mathcal{S}_G} - 2^{n-1} = 2^{n-1}$ different $P$'s in $\mathcal{S}_G$. 
    Hence, we have $A_{b,b'} = \delta_{b,b'}$.
    
    We then have $\mathbf{w} = \mathbf{Q}^{-1} \mathbf{a} = \mathbf{Q_0}^{T} \mathbf{a}$, that is,
    \begin{equation}
        \Tr(\rho P) = \sum_{b\in \mathbb{F}_2^n} (-1)^{f_b(P)} a_b.
        \label{eq:Tr=linear(a)}
    \end{equation}
    If we assume that $a_b > {1}/{2}$ for some $b$, we can deduce from Eq.~\eqref{eq:Tr=linear(a)} that
    \begin{equation}
        \mathrm{sgn}(\Tr(\rho P)) = (-1)^{f_{b}(P)}, \quad \forall P\in \mathcal{S}_G^+, 
    \end{equation}
    or alternatively, the $b$ term dominates the entire sum. Here, we note that picking $P=I^{\otimes n}$ in Eq.~\eqref{eq:Tr=linear(a)} is trivially the case $\Tr(\rho)=1$.
\end{proof}

\section{Proof of Theorem~\ref{thm:main}}\label{app:theorem-sample-complexity} 
Here we present the proof of Theorem~\ref{thm:main}, which we now recall.
\renewcommand{\thetheorem}{\ref{thm:main}}
\begin{theorem}
    With probability no less than $1-\delta$, the vector $\mathbf{a}$ corresponding to graph-diagonal elements of $\rho$ can be learned such that $\Vert \Delta\mathbf{a} \Vert_2 \leq \varepsilon$ using $\mathcal{O}\!\left((n + \log(1/\delta))/{\varepsilon^4}\right)$ samples.
\end{theorem}
\renewcommand{\thetheorem}{\arabic{theorem}} 

\begin{proof}
We reiterate that the shadow tomography protocol used here relies on the fact that we can simultaneously measure $P\otimes P$ (and consequently the absolute values $\abs{\Tr(\rho P)}$) upon performing $n$-fold two-copy Bell basis measurements: see the Supplementary Material of Ref.~\cite{huang2021information} for more details. 
Unlike the general two-copy shadow tomography protocol (i.e., Bell sampling) in~\cite{huang2021information}, our protocol does not require an additional majority vote due to Theorem~\ref{thm:signs}.

For any $P_i\in \mathcal{S}_G^+$, we have
\begin{equation}
    c_i \equiv \abs{\Tr(\rho P_i)}^2
    = \Tr((\rho\otimes\rho)(P_i\otimes P_i))
    = (+1)\Pr[\text{Eig}(P_i\otimes P_i) = +1] + (-1)\Pr[\text{Eig}(P_i\otimes P_i) = -1].
\end{equation}
Let
\begin{equation}
     \ket{\phi_{\beta}} \equiv \bigotimes_{k=1}^n \left(Z^{\beta[2k-1]}X^{\beta[2k]} \otimes I \ket{\Phi_{00}}\right),
\end{equation}
where we have the Bell state $\ket{\Phi_{00}} \equiv (\ket{00}+\ket{11})/\sqrt{2}$, $\beta$ is a bitstring of length $2n$, and $\beta[2k]$ is the $2k$-th bit in $\beta$ corresponding to node $k$. 
With Bell sampling, we find that
\begin{align}
    c_i
    &= \quad \mathop{\mathbb{E}}_{\beta}\left[\mel{\phi_{\beta}}{P_i\otimes P_i}{\phi_{\beta}}_{\mathrm{random}}\right],\\
    &= \sum_{\beta\in \{0,1\}^{2n}} \Pr(\beta)\mel{\phi_{\beta}}{P_i\otimes P_i}{\phi_{\beta}},\\
    &=\,(+1)\!\!\!\!\sum_{\beta\in \{0,1\}^{2n}}\!\!\!\!\Pr(\beta)\Pr(\prod_{k=1}^n \text{Eig}(\sigma_k^{(i)}\otimes\sigma_k^{(i)})=+1\Bigg|\,\beta)+(-1)\!\!\!\!\sum_{\beta\in \{0,1\}^{2n}}\!\!\!\!\Pr(\beta)\Pr(\prod_{k=1}^n \text{Eig}(\sigma_k^{(i)}\otimes\sigma_k^{(i)})=-1\Bigg|\,\beta),
\end{align}
where $P_i = \bigotimes_{k=1}^n \sigma_k^{(i)}$, and $\Pr(\beta)$ is the probability that the outcome is the bitstring $\beta$ (given a discrete probability distribution over the space of all possible bitstrings $\beta$, which depends on $\rho$). 
Let $N$ be the total number of identical pairs $\rho^{\otimes 2}$ available for the entire Bell measurement experiment. 
Then, the empirical quantity $\hat{c}_i\equiv\hat{c}_i(\mathcal{B})$ reads
\begin{equation}
    \hat{c}_i = \frac{1}{N}\sum_{j=1}^N\mel{\phi_{\beta}}{P_i\otimes P_i}{\phi_{\beta}}_{j,\mathrm{random}}= \max\left(0,\,\,\frac{1}{N}\sum_{\beta\in {\mathcal{B}}}\mel{\phi_{\beta}}{P_i\otimes P_i}{\phi_{\beta}}\right),
\end{equation}
where ${\mathcal{B}}$ is the multiset of bitstrings observed in the experiment, with $\abs{\mathcal{B}} = N$. 
Notice that the empirical frequency of $\beta$ is already embedded in the above summation. 
To clarify, let $m_{\beta} \equiv \Pr(\beta)$, so that we have the empirical frequency $\hat{m}_{\beta} \equiv \abs{\{x\in\mathcal{B}:x=\beta\}}/N$. 
This means that
\begin{equation}
    \hat{c}_i = \max\left(0,\,\,\sum_{\beta\in \{0,1\}^{2n}} \hat{m}_{\beta}\mel{\phi_{\beta}}{P_i\otimes P_i}{\phi_{\beta}}\right)
\end{equation}
where $\mathbb{E}[\hat{c}_i] = c_i$. Via Hoeffding's inequality \cite{hoeffding1963prob}, we obtain the concentration bound
\begin{equation}
    \Pr( \abs{\hat{c}_i-\mathbb{E}[\hat{c}_i]}\geq \varepsilon)=\Pr( \abs{\hat{c}_i-c_i}\geq \varepsilon) \leq 2\exp(-\frac{1}{2}N\varepsilon^2)
\end{equation}
for every $i$. Here, we again state that all $\beta$ are drawn from a fixed discrete distribution.

Now, suppose that we have
\begin{equation}
    \abs{\hat{c}_i - c_i} \leq \varepsilon^2
\end{equation}
for any $P_i\in \mathcal{S}_G^+$. Then, using the fact that $\sqrt{x+y}\leq \sqrt{x}+\sqrt{y}$, we obtain
\begin{equation}\label{eqn:bounds-c-to-sqrtc}
    \max(0,\sqrt{c_i} -\varepsilon) \leq \sqrt{\hat{c}_i} \leq \sqrt{c_i} + \varepsilon \quad \Longrightarrow \quad \abs{\sqrt{\hat{c}_i} - \sqrt{c_i}} = \abs{\hat{w}_i - w_i} = \abs{\Delta w_i}\leq \varepsilon.
\end{equation}
Given that $ \Delta\mathbf{a}=\mathbf{Q} \Delta\mathbf{w}$, we can, therefore, bound the $2$-norm $\Vert \Delta \mathbf{a} \Vert_2$ with
\begin{equation}
    \Vert \Delta \mathbf{a} \Vert_2 = \Vert \mathbf{Q}\Delta \mathbf{w} \Vert_2 = \sqrt{\langle\mathbf{Q}^T\mathbf{Q}\Delta\mathbf{w},\Delta\mathbf{w}\rangle}= \sqrt{\bigg\langle\mathbf{Q_0}^{-1}\frac{1}{2^n}\mathbf{Q_0}\Delta\mathbf{w},\Delta\mathbf{w}\bigg\rangle} = \frac{1}{\sqrt{2^n}}\Vert \Delta \mathbf{w} \Vert_2 \leq  \frac{1}{\sqrt{2^n}}\Vert \varepsilon\mathbf{1}_{2^n} \Vert_2 = \varepsilon,
\end{equation}
where $\mathbf{Q}^T=\mathbf{Q_0}^{-1}$, and $\mathbf{1}_{2^n}$ is a column vector of $2^n$ ones. Hence, we see that
\begin{equation}
    \Pr(\bigcap_{i=0}^{2^n-1}\{\abs{\hat{c}_i - c_i} \leq \varepsilon^2\}) \leq \Pr(\Vert \Delta \mathbf{a} \Vert_2 \leq \varepsilon). 
\end{equation}
Now,
\begin{equation}
    \Pr(\bigcap_{i=0}^{2^n-1}\{\abs{\hat{c}_i - c_i} \leq \varepsilon^2\}) \geq  1 - 2^{n+1}\exp(-\frac{1}{2}N\varepsilon^4),
\end{equation}
where we use the intersection bound for probabilities $\Pr(\bigcap_{j} E_j) \geq 1 - \sum_j \delta_j$ for any event $E_j$ that satisfies $\Pr(E_j) \geq 1-\delta_j$, and we assume $\exp(-N\varepsilon^4/2)\leq 2^{-(n+1)}$. Therefore,
\begin{equation}
    \Pr(\Vert \Delta \mathbf{a} \Vert_2 \geq {\varepsilon}) \leq 2^{n+1}\exp(-\frac{1}{2}N\varepsilon^4).
\end{equation}
Following the above, we see that $\Vert \Delta \mathbf{a} \Vert_2 \leq \varepsilon$ can be achieved with some high probability $1-\delta$ whenever
\begin{equation}
    N = \mathcal{O}\!\left(\frac{\log(2^n/\delta)}{\varepsilon^4}\right) = \mathcal{O}\!\left(\frac{n+\log(1/\delta)}{\varepsilon^4}\right).
\end{equation}
Finally, we can also bound the estimate for fidelity $F$ with respect to $\ket{\Phi_G}$ since $\abs{\Delta F}\leq \Vert \Delta \mathbf{a} \Vert_{\infty} \leq \Vert \Delta \mathbf{a} \Vert_2$. 
\end{proof}

\section{
Proof of Theorem~\ref{thm:fidelity_random}:
Estimating fidelity via random sampling \label{app:random_sampling}}

In this appendix, we give the proof of Theorem~\ref{thm:fidelity_random} which gives the sample complexity to estimate the fidelity with random sampling, which we recall for convenience of the reader:
\renewcommand{\thetheorem}{\ref{thm:fidelity_random}}
\begin{theorem}
    Let $M = \mathcal{O}(\log(1/\delta)/\varepsilon^2)$ be the number of randomly selected stabilizer measurement settings, independent of the size of the graph. 
    With probability no less than $1-2\delta$, an element of $\mathbf{a}$ can be learned with error $2\varepsilon$ using $\mathcal{O}(\log(1/\varepsilon^2\delta)/\varepsilon^4)$ samples.
\end{theorem}
\renewcommand{\thetheorem}{\arabic{theorem}} 

\begin{proof}
To begin with the proof, we first consider the special case of estimating fidelity 
(that is, the largest element in $\mathbf{a}$) 
and then generalize the result to any element of $\mathbf{a}$. We start with the expression for the true fidelity:
\begin{equation}
    F = \frac{1}{2^n}\sum_{i=0}^{2^n -1} \sqrt{c_i},
\end{equation}
in which $c_i \equiv \abs{\Tr(\rho P_i)}^2$ and $P_i \in \mathcal{S}_G^+$.
Notice that the classical post-processing scales unfavorably with the number of qubits $n$. 
We can reduce the complexity of this post-processing by only considering $M$ randomly sampled $\sqrt{c_i}$'s (these $c_i$'s can repeat), where $M\ll 2^n$, similar to the investigation in Ref.~\cite{flammia2011direct}. 

We construct a variable $\mathcal{F}$ which selects over the values $\sqrt{c_i}$ uniformly at random, that is,
\begin{equation}\label{eq:FF}
    \mathcal{F} \equiv \left\{ \sqrt{c_i}\,\, , \,\,\frac{1}{2^n}\right\}_{i=0}^{2^n -1}
\end{equation}
which means that
\begin{equation}
    \Pr(\mathcal{F} = \sqrt{c_i}) = \frac{1}{2^n}, \quad i \in \{0, 1, \ldots, 2^n - 1\},
\end{equation}
The expected value of $\mathcal{F}$ is then given by
\begin{equation}
    \mathbb{E}[\mathcal{F}] = \frac{1}{2^n} \sum_{i=0}^{2^n - 1} \sqrt{c_i} = F.
\end{equation}
We draw \(M\) independent samples \(\mathcal{F}_1,\dots,\mathcal{F}_M\) of the random variable \(\mathcal{F}\) defined in Eq.~\eqref{eq:FF}, and define the random variable
\begin{equation}
    Y_M \equiv \frac{1}{M}\sum_{j=1}^M \mathcal{F}_j,
\end{equation}
where \(\mathbb{E}[Y_M]=F\) by construction.
For each sample $\mathcal{F}_j$, we independently pick an index
\begin{equation}
    I_j \sim \operatorname{Unif}\{0,1,\dots,2^n-1\},
\end{equation}
and set \(\mathcal{F}_j=\sqrt{c_{I_j}}\).
We note that the samples are drawn with replacement, so indices \(I_j\) may repeat.
For reference,
let $\mathcal{S}_M=\{I_1,\dots,I_M\}$ be the sequence (multiset) of selected indices.
Now we can consider an unbiased estimator to the fidelity $F$
\begin{equation}
    \widetilde{Y}_M \equiv \frac{1}{M}\sum_{s\in \mathcal{S}_M}\sqrt{c_s}\ \ ,
\end{equation}
By Hoeffding's inequality, 
\begin{equation}
    \Pr(\abs{\smash[t]{\widetilde{Y}_M} - \mathbb{E}[Y_M]}\geq \varepsilon') =\Pr(\abs{\smash[t]{\widetilde{Y}_M} - F}\geq \varepsilon') \leq 2\exp(-2M(\varepsilon')^2) \equiv \delta',
\end{equation}
where $0\leq \sqrt{c_j} \leq 1$ for any $j$, and $\varepsilon'$ is an additive error. 
Therefore, to ensure $\abs{\smash[t]{\widetilde{Y}_M} - F}\leq \varepsilon'$ with success probability no less than $1-\delta'$, the required number of randomly sampled $\mathcal{F}$ is
\begin{equation}
    M = \mathcal{O}\!\left(\frac{\log(1/\delta')}{(\varepsilon')^2}\right).
    \label{eq:thm3_M}
\end{equation}
We then approximate this $\widetilde{Y}_M$ via a Bell sampling experiment given $N$ samples. From the experiment, we obtain the estimates $\sqrt{\hat{c}_s}$ to $\sqrt{c_s}$, giving us 
\begin{equation}
    \hat{Y}_M \equiv \frac{1}{M}\sum_{s \in \mathcal{S}_M}\sqrt{\hat{c}_s}\, .
\end{equation}
Following the steps that led to Ineq.~\eqref{eqn:bounds-c-to-sqrtc}, we have
\begin{equation}
    \abs{\smash[t]{\hat{Y}_M - \widetilde{Y}_M}} \leq \frac{1}{M}\sum_{s\in \mathcal{S}_M}\abs{\sqrt{\hat{c}_s} - \sqrt{c_s}} \leq \sqrt{\varepsilon''},
\end{equation}
and so
\begin{equation}
    \Pr(\bigcap_{s\in \mathcal{S}_M}\{\abs{\hat{c}_s - c_s} \leq \varepsilon''\}) \leq \Pr(\abs{\smash[t]{\hat{Y}_M - \widetilde{Y}_M}} \leq \sqrt{\varepsilon''}),
\end{equation}
for another additive error $\varepsilon''$. Hence, we have the concentration bound
\begin{equation}
    \Pr(\abs{\smash[t]{\hat{Y}_M - \widetilde{Y}_M}} \geq \varepsilon'') \leq 2M\exp(-\frac{1}{2}N(\varepsilon'')^4) \equiv \delta''.
\end{equation}
We can then conclude that the sample complexity of the Bell sampling experiment should scale like the following:
\begin{equation}
    N = \mathcal{O}\!\left(\frac{\log(M/\delta'')}{(\varepsilon'')^4}\right).
    \label{eq:thm3_N}
\end{equation}
Finally, following the triangle inequality and the previous bounds, we obtain
\begin{equation}
    \abs{\smash[t]{\hat{Y}_M -F}} \leq \abs{\smash[t]{\hat{Y}_M - \widetilde{Y}_M}} + \abs{\smash[t]{\widetilde{Y}_M - F}} \leq \varepsilon'' + \varepsilon'.
\end{equation}
Moreover, by using the intersection bound for probabilities, we have the concentration bound 
\begin{equation}
    \Pr(\abs{\smash[t]{\hat{Y}_M -F}} \geq \varepsilon) \leq \delta' + \delta''.
    \label{eq:thm3_concentration_bound}
\end{equation}
Therefore, by combining Eqs.~\eqref{eq:thm3_N}, \eqref{eq:thm3_M}, and \eqref{eq:thm3_concentration_bound},
we find that the fidelity $F$ can be estimated by the estimator $\smash[t]{\hat{Y}_M}$ with an error $\varepsilon'+\varepsilon''$ and success probability of at least $1-(\delta'+\delta'')$ by requiring
 \begin{equation}    
    N = \mathcal{O}\!\left(
    \frac{\log\left(\frac{\log(1/\delta')}{\delta'' \varepsilon'^2}\right)}{(\varepsilon'')^4}\right)
\end{equation}
samples.
The advantage of using this approach is evident from the fact that both $N$ and $M$ are independent of the graph order $n$. 
Setting $\varepsilon'=\varepsilon''$ and $\delta' = \delta''$ gives us the promised bound. 
Here, we note that the function $\log(x)$ dominates $\log\log(x)$ for large $x$.

It is possible to extend this proof to the remaining diagonal elements $a_b$. 
Assuming that $\mel{\Phi_G}{\rho}{\Phi_G}> 1/2$, we can write
\begin{equation}
    a_b = \frac{1}{2^n}\sum_{i=0}^{2^n -1} (-1)^{f_b(P_i)} \abs{\Tr(\rho P_i)} = \frac{1}{2^n}\sum_{i=0}^{2^n -1} (-1)^{f_b(P_i)} \sqrt{c_i}
\end{equation}
following Theorem~\ref{thm:signs}. 
We then construct a variable $\mathcal{A}_b$ which selects over values uniformly at random, so that $\mathbb{E}[\mathcal{\mathcal{A}}_b] = a_b$ over a random choice of $P_i$:
\begin{equation}
    \mathcal{A}_b \equiv \left\{ (-1)^{f_b(P_i)}\sqrt{c_i}\,\, , \,\,\frac{1}{2^n}\right\}_{i=0}^{2^n -1}.
\end{equation}
which means that
\begin{equation}
\Pr\left(\mathcal{A}_b =(-1)^{f_b(P_i)}\sqrt{c_i}\right) = \frac{1}{2^n}, \quad i \in \{0, 1, \ldots, 2^n - 1\}.
\end{equation}
Similarly as before,
we draw $M$ independent samples of the random variable $\mathcal{A}_b$,
and define the random variable
\begin{equation}
    Y_M^b\equiv \frac{1}{M}\sum_{j=1}^M \mathcal{A}_{b,j}\,\, ,
\end{equation}
where $\mathbb{E}[Y_M^b]=a_b$, and all $\mathcal{A}_{b,j}$'s are mutually independent. 
Let $\mathcal{S}_M=\{I_1,\dots,I_M\}$ be the sequence (multiset) of selected indices, and set 
$\mathcal{A}_{b,j}=(-1)^{f_b(P_{I_j})}\sqrt{c_{I_j}}$ where $I_j\in\mathcal{S}_M$.
This time, we consider an unbiased estimator of $a_b$,
\begin{equation}
    \widetilde{Y}_M^b \equiv \frac{1}{M}\sum_{s\in \mathcal{S}_M}(-1)^{f_b(P_s)}\sqrt{c_s}\,\, , 
\end{equation}
that we construct after random sampling. By Hoeffding's inequality, 
\begin{equation}
    \Pr(\abs{\smash[t]{\widetilde{Y}_M^b} - \mathbb{E}[Y_M^b]}\geq \varepsilon') =\Pr(\abs{\smash[t]{\widetilde{Y}_M^b} - a_b}\geq \varepsilon') \leq 2\exp(-\frac{1}{2}M(\varepsilon')^2) \equiv \delta''',
\end{equation}
since $-1\leq (-1)^{f_b(P_j)}\sqrt{c_j} \leq 1$ for any $j$. 
Hence, the complexity in the post-processing level is
\begin{equation}
    M = \mathcal{O}\!\left(\frac{\log(1/\delta''')}{(\varepsilon')^2}\right).
\end{equation}
The approximation of $\widetilde{Y}_M^b$ by a Bell sampling experiment given $N$ copies of $\rho^{\otimes 2}$ will be similar to before. 
This key result allows estimation via Bell sampling to be dimension-independent.
\end{proof}


\section{Direct graph-diagonal estimation (DGE)}
\label{app:dge_description}

Here, we describe the general procedure for the direct graph-diagonal estimation (DGE) protocol.
The core idea of DGE is to estimate the expectation values, $\Tr(\rho P)$, of each stabilizer element $P \in \mathcal{S}^+_{G}$ using only node-local (single-qubit) measurements on one copy of the state at a time.
A naive implementation would require a separate measurement for each of the $d=2^n$ stabilizer elements $P_b$.

We can make DGE more efficient by measuring multiple stabilizer elements simultaneously.
This can clearly be seen for elements that have disjoint support, e.g., $Y_1 Y_2$ and $Y_3 Y_4$.
This grouping means that allocating $k$ measurement shots to estimating each stabilizer element expectation value $\braket{P}$ requires fewer than $k \times d$ total shots, as we can combine some stabilizers.

We can optimize the measurement settings further by combining elements that have overlapping support, provided they share the same Pauli operator on the overlapping qubits.
For example, we can extract both $Y_1 Y_2$ and $Y_2 Y_3$ from a single $Y_1 \otimes Y_2 \otimes Y_3$ measurement setting.

We use this second optimization approach to obtain the DGE results presented in Figure~\ref{fig:norm_dF} and Figure~\ref{fig:N_complexity} of the main text.
While this grouping of multiple stabilizer elements into a single measurement setting still provides an unbiased estimator for the diagonal elements, we note that the variance of the estimator depends on how we combine the stabilizer elements.

\section{Numerical simulation procedure}
\label{app:simulation_procedure}

We now outline the simulation details used to obtain the numerical results discussed in Sec.~\ref{sec:results}.
The full code implementation is publicly available at \url{https://github.com/Naphann/bell-sampling-for-quantum-network}.
We performed all simulations using stabilizer circuit simulator, Stim~\cite{gidney-stim}.

We begin by defining the explicit stabilizer group ordering we utilized in the simulation.
We then describe the specific simulation methods used to obtain measurement samples for both the DGE and BSQN protocols.

\subsection{Stabilizer group ordering}

To connect the graph-diagonal vector $\mathbf{a}$ to the stabilizer expectation values, we first define a consistent ordering for the $2^n$ elements of the stabilizer group $\mathcal{S}_G^+$.
We assume a fixed ordering for the $n$ qubits (vertices) of the graph $G$, indexed from $j=1$ to $n$.

This indexing defines the $n$ stabilizer generators, $\{S_j\}_{j=1}^n$, where $S_j$ is the generator associated with the $j$-th qubit, $S_j = X_j \bigotimes_{k \in \text{NN}(j)} Z_k$.
We then map each $n$-bit string $b \in \mathbb{F}_2^n$ to a unique stabilizer element $P_b \in \mathcal{S}_G^+$ via the product of these generators:
\begin{equation}
    P_b \equiv \prod_{j=1}^n (S_j)^{b_j}.
\end{equation}
Under this convention, the bit string $b = 0^n = \overline{0 \ldots 0}$ maps to the identity element $P_{0^n} = I$.
Any string $b$ with a Hamming weight $\mathrm{wt}(b)=1$ and a single set bit at index $j$ maps to the $j$-th generator, $P_b = S_j$---for example, when $b = 00100\ldots0$, we would get $P_b = S_3$.

This ordering allows us to define a vector of stabilizer expectation values, $\mathbf{w}$, where each element $w_b = \Tr(\rho P_b)$ corresponds to the indexed stabilizer element $P_b$.
As established in Lemma~\ref{lem:exactrep}, the graph-diagonal vector $\mathbf{a}$ is calculated from these expectation values; for this specific ordering, the relationship is a Walsh-Hadamard transform, which allows us to compute $\mathbf{a}$ efficiently from $\mathbf{w}$.

\subsection{DGE simulation}

Our simulation focuses on complete graph states.
We first analyze the structure of the $d=2^n$ stabilizer elements $P_b$ for a complete graph.

First, any $P_b$ corresponding to a bit string $b$ with an odd Hamming weight has full support (i.e., it is a tensor product of $X$ and $Z$ operators).
There are $d/2$ such elements, and since their supports fully overlap and cannot be combined, we require $d/2$ distinct measurement settings (circuits) to estimate all $d/2$ expectation values.
Second, any $P_b$ where $b$ has an even Hamming weight has partial support, consisting of a tensor product of $Y$ and $I$ operators.
Crucially, this structure means we can estimate all $d/2$ of these even-weight elements from a single measurement setting (i.e., measuring all qubits in the $Y$ basis).

Therefore, our DGE implementation requires $d/2 + 1$ distinct measurement settings.
We allocate the total budget of $N_s$ copies (shots) among these settings.
If $N_s$ is not evenly divisible by $(d/2 + 1)$, we randomly assign the remaining shots, one by one, to the settings until the budget is used up.

We also note a key optimization in our simulation.
Instead of simulating $(d/2 + 1)$ distinct circuits, we simulate a single noisy circuit and track the noise ($Z$ terms) affecting each qubit.
We then flip the measurement results accordingly to efficiently compute the parity for all $d/2+1$ settings.
This method is computationally equivalent to running $(d/2 + 1)$ separate circuits but is significantly more efficient for simulation.

\subsection{BSQN simulation}

For the BSQN protocol, we simulated the two-copy pairwise Bell measurement as described in the main text.
The measurement outcomes, corresponding to the four Bell states, were recorded and post-processed inferring the three Pauli operators (i.e., $XX$, $YY$, and $ZZ$) to simultaneously estimate the magnitudes, $|\Tr(\rho P_b)|^2$, for all $d=2^n$ stabilizer elements.

\section{Protocol independence from graph topology}
\label{app:graph-independence}

We describe here why our estimation protocols are independent of the underlying graph topology.
The graph-state basis $\mathcal{G} = \{ \ket{\Phi_b} \}$ is, by definition, specific to a particular graph state $\ket{\Phi_G}$.
A state $\rho$ that is diagonal in one graph's basis will not, in general, be diagonal in the basis of a different graph.
We show, however, that if two different noisy graph states (originating from different graph topologies) possess the same logical error distribution $\mathbf{a}$ in their respective graph-state bases, our estimation protocols will behave identically for both.

This independence comes from our protocol's methodology, from estimating the expectation values over all stabilizer elements and the post-processing method.
As discussed in Sec.~\ref{sec:errordetection}, the effect of a Pauli channel on $\ket{\Phi_G}$ is to produce a diagonal state $\rho = \sum_b a_b \ketbra{\Phi_b}{\Phi_b}$, where each $a_b$ corresponds to the probability of a logical error $U_b = \bigotimes_i Z_i^{b_i}$ occurring.
One would detect this logical error $U_b$ by looking at the expectation values of a set of stabilizer elements $\{P_c\}$, where a measurable effect (a syndrome) is produced only when $U_b$ and $P_c$ anti-commute, which can be expressed as $\mathrm{wt}(b \land c) \pmod 2 = 1$.
Essentially, because both the noise channels and the graph-diagonal basis are defined in this related way, this justifies using a single topology (in our case, a complete graph) as a representative for the protocol's numerical simulation.

\section{Errors of estimation with a different implementation of DGE}
\label{app:errors-with-different-dge}

\begin{figure}[t]
    \centering
    \includegraphics[width=1\linewidth]{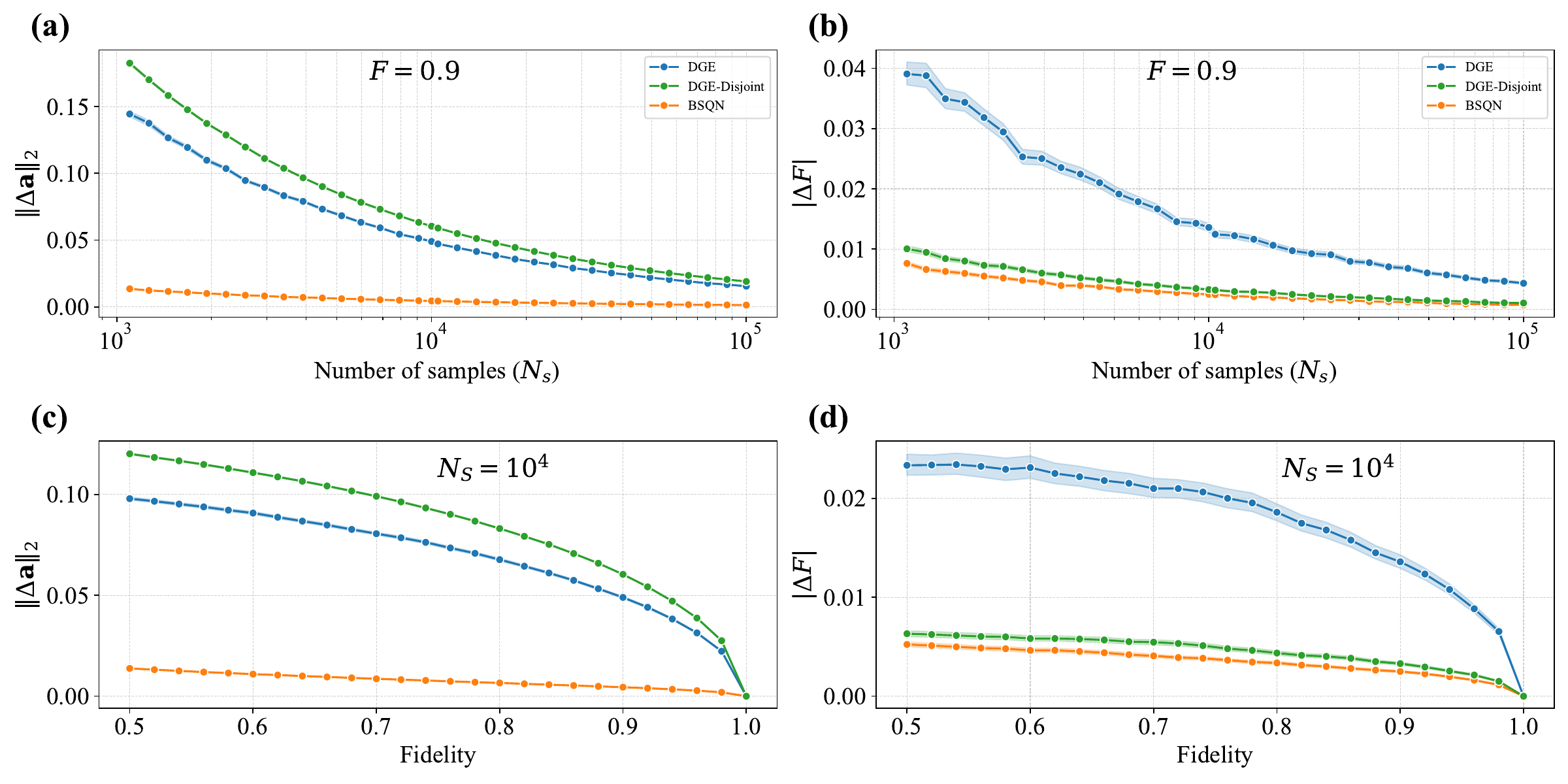}
    \caption{Performance comparison of BSQN, DGE, and DGE-Disjoint (this appendix) for an $8$-qubit complete graph state. The DGE-Disjoint protocol pairs disjoint even-weight stabilizers. This improves fidelity estimation (b, d) but increases the $2$-norm error (a, c), showing the implementation tradeoff compared to the DGE presented in the main text.} 
    \label{fig:norm_dF-appendix}
\end{figure}

As we mentioned in Appendix~\ref{app:dge_description}, one can implement the DGE protocol in multiple ways by combining stabilizer elements differently.
Here, we analyze an alternative implementation, ``DGE-Disjoint,'' where we only combine stabilizer elements that have disjoint support.
To be specific, for a complete graph with an even number of qubits ($n=8$), we analyze the stabilizer elements $P_b$ with even Hamming weight.
For any such element (e.g., $P_b = Y_1 Y_3 Y_5 Y_8$), there is always a unique partner $P_c$ with disjoint support ($P_c = Y_2 Y_4 Y_6 Y_7$) such that their product is all-$Y$  with full support, $P_b \otimes P_c = \bigotimes_{j=1}^n Y_j$.
This pairing strategy is optimal in the sense that it utilizes the measurement results from all $n$ qubits in the $Y$-basis for every shot.

Figure~\ref{fig:norm_dF-appendix} shows the estimation errors for this DGE-Disjoint implementation, using the same simulation parameters (1000 trials per point) as in Figure~\ref{fig:norm_dF} presented in the main text.
Note that we vary $N_s$ from $10^3$ to $10^5$ in this comparison, a higher starting point than in Figure~\ref{fig:norm_dF}, to accommodate the higher minimum sample requirement of DGE-Disjoint.
The results demonstrate a tradeoff.
This DGE-Disjoint method shows significantly improved fidelity estimation compared to the DGE (with overlapped stabilizer elements) method, making the estimates closer to BSQN, although BSQN still performs slightly better.
However, this comes at the cost of a larger $2$-norm error indicating that it performs worse for estimating the diagonal vector.

We conjecture that the larger $2$-norm errors stem from having fewer samples allocated to each stabilizer element compared to the DGE approach from the main text.
The improved fidelity estimation is less clear, but we conjecture that it relates to non-zero covariance, $\mathrm{Cov}(w_b, w_c) \neq 0$, introduced by the measurement grouping.
These correlations may accumulate for fidelity (an all-positive sum), whereas they might partially cancel for other diagonal elements (which have mixed-sign sums).
A rigorous analysis of the covariance matrix for each implementation is required to confirm this hypothesis, which we leave for future work.


\twocolumngrid

\bibliography{reference}

\end{document}